%% file: main.tex
\documentclass{article}
\usepackage{fullpage}
\usepackage{parskip}
\usepackage{standalone}
\usepackage{graphicx}
\usepackage{booktabs}
\usepackage{subcaption}
\usepackage{hyperref}
\usepackage{amsmath}
\usepackage{amsfonts}
\usepackage[table]{xcolor}
\usepackage{xfrac}
\usepackage{tikz}
\usetikzlibrary{arrows}
\usetikzlibrary{decorations.pathreplacing}
\usetikzlibrary{calc}
\usetikzlibrary{shapes.geometric}
\usetikzlibrary{decorations.markings}
\usetikzlibrary{shapes.geometric}
\usetikzlibrary{patterns,snakes}
\usetikzlibrary{decorations.text}
\usetikzlibrary{arrows.meta}

\title{Discrete Event Population Updates: finding game theoretic
emergent behaviour in queueing systems with simulation}
\author{Vincent Knight, Geraint I. Palmer-Liyu \& Thomas Hutton}
\date{}

\begin{document}
\maketitle

\begin{abstract}
Strategic behaviour in queueing systems has been studied extensively in the
behavioural queueing literature, but almost exclusively for systems that
admit closed-form expressions for the cost or utility experienced by a
strategic user. Evolutionary game theory offers a mature framework for
analysing populations whose individual payoffs depend on the composition of
the population itself, and would in principle apply to a much wider class
of queueing systems; its application has, however, been constrained by the
same closed-form requirement. We introduce Discrete Event Population
Updates (DEPU), a general algorithmic framework that couples a single long
run of a discrete event simulation (DES) directly to an evolutionary population
update rule, removing that constraint. We present two implementations:
Discrete Event Replicator Dynamics (DERD), which follows an Euler
discretisation of the replicator dynamics equation, and Discrete Event
Moran Replacement (DEMR), which maintains a finite population updated via
Moran-style copying events. Both are 
 applied to a multi-server
jockeying model for which no closed-form fitness expressions are
available. On the jockeying model considered, DEPU reaches comparable
precision tens of times faster than the standard practice of nesting short
simulations inside an outer evolutionary loop, and because each operating
point then costs only a single simulation run it also makes systematic
parameter sweeps tractable. This brings the toolkit of
evolutionary dynamics within reach of any system a modeller can build in
a discrete event simulator.
\end{abstract}

\section{Introduction}

Most real queueing systems admit no closed-form expression for the cost or
utility experienced by a strategic user, yet the behavioural queueing
literature is built almost entirely on the exceptions. Naor's seminal study
of the observable M/M/1 queue~\cite{naor1969regulation} identified an
equilibrium threshold strategy for individuals choosing whether to join a
queue upon observing it, and~\cite{bell1983individual} characterised
equilibrium arrival rates in the unobservable multi-server setting; both
rely on tractable closed-form expressions for the per-customer expected
cost. Comprehensive treatments of this analytical line are given
in~\cite{hassin2016rational, hassin2003queue}. Outside this narrow class
of analytically tractable systems, strategic queueing has remained largely
out of reach.

Evolutionary game theory~\cite{hofbauersigmund1998, maynardsmith1982,
nowak2006evolutionary, taylorjonker1978} provides a mature framework for
analysing populations whose individual payoffs depend on the decisions made by
all individuals in the 
population itself, and would in principle apply to any queueing system in which
the effectiveness of a strategy depends on what others are doing. Consider, for
example, the problem of finding a seat in a multi-storey library: some
individuals start at the top floor and work their way down, others start at the
bottom and work upward. The effectiveness of either strategy depends on its
prevalence in the population, and the natural equilibrium concept is an
evolutionary one. Yet the closed-form fitness expressions that evolutionary
dynamics has classically required are unavailable here, as they are for almost
every queueing system of practical interest.

In this paper we introduce \emph{Discrete Event Population Updates}
(DEPU), a general algorithmic framework that removes this constraint.
DEPU couples a single long run of a DES~\cite{pidd98, robinson14}
directly to an evolutionary
population update rule, rather than nesting repeated short simulations
inside an outer optimisation or evolutionary loop. The single-run
architecture is what unlocks the framework: each customer-departure event
observed in the simulator contributes to a running fitness estimate for
the strategy that customer was using, and the population vector evolves in
lockstep with the simulation clock. We present two implementations,
Discrete Event Replicator Dynamics (DERD), which follows an Euler
discretisation of the replicator dynamics equation, and Discrete Event
Moran Replacement (DEMR), which maintains a finite population updated via
Moran-style copying events. Both are applied to a multi-server
jockeying model (Figure~\ref{fig:jockeying}), a generalised form
of~\emph{jockeying}~\cite{haight1958} in which individuals choose an order
in which to seek service from a selection of queues. The library example
above is one instance of this model.

The contributions of this paper are as follows:

\begin{itemize}
    \item \textbf{Discrete Event Population Updates.} A general algorithmic
        framework that brings the toolkit of evolutionary dynamics within
        reach of any strategic system that can be modelled in a discrete
        event simulator. DEPU is decoupled into a fitness-estimation step
        and a population-update step, so that any revision protocol
        expressible as a map from a fitness vector to a population update
        can be substituted in.
    \item \textbf{Two reference implementations.} Discrete Event Replicator
        Dynamics (DERD), which follows an Euler discretisation of the
        replicator dynamics equation, and Discrete Event Moran
        Replacement (DEMR), which maintains a finite discrete population
        updated via Moran-style copying and removal events.
    \item \textbf{Computational efficiency.} On the jockeying model
        considered here, DEPU reaches comparable precision tens of times
        faster than the standard practice of nesting short simulations
        inside an outer evolutionary loop. Because each operating point
        then costs only a single DES run, DEPU also makes systematic
        parameter sweeps tractable.
    \item \textbf{A jockeying queueing model.} A concrete instance of a
        strategic queueing problem with no closed-form fitness, modelling
        individuals choosing how to search through a collection of
        multi-server queues, together with an analysis of its emergent
        equilibrium behaviour under DEPU.
\end{itemize}

The paper is organised as follows. Section~\ref{sec:literature_review}
reviews the behavioural queueing and evolutionary game theory literature on
which we build. Section~\ref{sec:general_model} introduces the general
strategic jockeying model that serves as our running example.
Section~\ref{sec:baseline} applies classical replicator dynamics and the
Moran process to this model using DES to estimate fitness, establishing a
baseline and exposing the computational cost of the nested-loop approach.
Section~\ref{sec:depu} introduces DEPU together with its two
implementations, validates them against Naor's observable queue, and applies
them to richer instances of the jockeying model.
Section~\ref{sec:performance} quantifies the speed-up of DEPU over the
baseline, and Section~\ref{sec:conclusion} concludes.

\section{Literature review}
\label{sec:literature_review}

\subsection{Background on behavioural queueing theory}

The study of strategic behaviour in queues was pioneered
in~\cite{naor1969regulation}, where it was shown that, in an observable M/M/1
queue where arriving customers can see the queue length before deciding to
join, the individually optimal strategy takes the form of a threshold rule:
join if and only if the observed queue length does not exceed some threshold
\(\tilde{n}\). The seminal work in~\cite{naor1969regulation} derives a
closed-form expression for this equilibrium threshold and also characterises
the socially optimal threshold, which is strictly lower. The gap between
individual and social optima reflects the negative externality that each
joining customer imposes on those already waiting: a phenomenon quantified as
the \emph{price of anarchy}~\cite{haviv2007poa, knight2017measuring,
roughgarden2002selfish}.

The distinction between observable and unobservable queues is fundamental in
behavioural queueing theory~\cite{shone2013comparisons}. In unobservable
systems, customers must decide whether to join without knowledge of the
current queue state. Congestion tolls for unobservable M/M/1 queues were
studied in~\cite{edelson1975congestion}, characterising socially optimal
joining rates as a companion to the observable analysis
of~\cite{naor1969regulation}. This line of work was extended to the
multi-server setting in~\cite{bell1983individual}, characterising the
equilibrium joining probability as a symmetric mixed Nash equilibrium.
Rational customer abandonments in unobservable queues, which are closely
related to the jockeying limits of the model considered in this paper, are
studied in~\cite{mandelbaum2000rational}. Comprehensive surveys across
observable and unobservable models are given in~\cite{hassin2016rational,
hassin2003queue}.

Strategic behaviour becomes considerably more complex in multi-server and
multi-facility settings. Optimal routing across a network of heterogeneous
service facilities is studied in~\cite{shone2020conservative}, formulating
the problem as a Markov decision process and proposing a conservative index
heuristic to manage its computational complexity. A containment result is
obtained in~\cite{shone2016containment} for multiple-facility Markovian
queueing systems, showing that individually optimal joining sets are always
subsets of the socially optimal joining sets; that is, individual rationality
leads to greater congestion than the social optimum. The question of which
queue to join is non-trivial even in simple parallel-server settings:
\cite{whitt1986deciding} demonstrates that joining the shortest queue is not
always individually optimal under general service distributions. In the
unobservable case, \cite{knight2013selfish} studies selfish routing in a
public service system and similarly finds that individually rational
behaviour increases congestion relative to the social optimum.

Applications to healthcare scheduling have motivated several developments in
behavioural queueing theory. Routing decisions in a network of intensive care
units are modelled in~\cite{knight2017measuring}, using the price of anarchy
as a measure of the cost of individual rationality relative to coordinated
management. Strategic interactions at the emergency medical service to
emergency department interface are modelled
in~\cite{panayides2023game} using asymmetric replicator dynamics,
demonstrating the applicability of evolutionary game-theoretic models to
healthcare queueing problems.

A further strand of related work concerns \emph{jockeying}: the behaviour of
customers who switch from one queue to another when a shorter one becomes
available~\cite{koenigsberg1966jockeying}. Early work in~\cite{haight1958}
provided analytical results on queue-length distributions in two-queue systems
with jockeying; this was extended in~\cite{disney1970jockeying} to handle
instantaneous jockeying with general customer selection policies. The present
paper extends the notion of jockeying to a strategic setting in which
individuals choose in advance an ordered search strategy for obtaining service
from a set of parallel nodes. This formulation does not admit closed-form
analysis, and therefore motivates the simulation-based approach developed here.

\subsection{Evolutionary Game theoretic models of emergent behaviour}\label{sec:rd_and_moran}

Evolutionary game theory provides a framework for analysing the emergence of
collective behaviour in populations of strategic agents, without requiring the
assumption that all individuals simultaneously solve for a Nash equilibrium.
Introduced in a biological context in~\cite{maynardsmith1982}, evolutionary
game theory replaces the static equilibrium concept with dynamic processes
that describe how strategies spread or contract within a population based on
their relative fitness. A central solution concept for infinite populations
is the \emph{evolutionarily stable strategy} (ESS): a strategy that, once
adopted by the whole population, cannot be invaded by a rare mutant using an
alternative strategy, and towards which a population is likely to gravitate.
For finite populations of size $M$ the analogous notion is the \emph{fixation
probability}: the probability that a single mutant of one type introduced
into a population of another type eventually takes over entirely. A strategy
is said to be selected for when its fixation probability exceeds \(1/M\),
the neutral drift baseline. These two concepts are asymptotically consistent:
the Moran process converges to the replicator dynamics equation as \(M \to
\infty\)~\cite{traulsen2005coevolutionary}.

Throughout this work we consider a finite \emph{strategy space} \(S\), where
each element \(s \in S\) represents a behaviour an individual may adopt. The
composition of the population is described either as a vector of proportions
\(x = (x_s)_{s \in S}\) for an infinite population, or as a vector of integer
counts \(v = (v_s)_{s \in S}\) for a finite population of size \(M = \sum_s
v_s\). The \emph{fitness} \(f_s\) of strategy \(s\) is a function of the
population composition: it measures how well an individual using strategy
\(s\) does when the rest of the population is distributed accordingly. The
question studied by evolutionary dynamics is then how the population
composition changes over time as a function of these fitnesses.

This work draws on two specific models of evolutionary dynamics that are
well-studied in the literature~\cite{hofbauersigmund1998,
nowak2006evolutionary}: the replicator dynamics (RD) equation and the
Moran process. The RD equation models the evolution of an
infinite population in continuous time; the Moran process models the
evolution of a finite population in discrete time. These are not the only
evolutionary dynamics, but they are the two canonical models for the
infinite and finite population regimes respectively. Other dynamics that
could readily be used with the proposed framework are discussed in
Section~\ref{sec:conclusion}.

\subsubsection{Replicator dynamics}\label{sec:replicator_dynamics}

The first model of emergent behaviour on which this work is based is the
RD equation:

\begin{equation}
    \frac{dx_s}{dt} = x_{s}(f_s(x) - \phi)
    \label{eqn:replicator_dynamics_equation}
\end{equation}

where \(\phi\) is the average fitness of a population
\(x\) given by:

\begin{equation}
    \phi = \sum_{s \in S}x_s f_s
\end{equation}

A stable population vector \(x\) is one that gives \(\frac{dx_s}{dt}=0\) for all
\(s\in S\) which implies that either:

\begin{itemize}
    \item \(x_s=0\): strategy \(s\) is not present in the population or,
    \item \(f_s=\phi\): the fitness of strategy \(s\) is equal to the average
        fitness.
\end{itemize}

This implies that at a stable population, no individual has an incentive to
modify their behaviour.

\subsubsection{Moran process}\label{sec:moran_process}
The second model of emergent behaviour considered here is the Moran
process~\cite{moran1958random}. This model is a stochastic process on a
population of constant size \(M\). The population is described by a vector
\(v\in\mathbb{Z}^{|S|}\), where \(v_i\) corresponds to the number of
individuals of type \(i\) (using \(s_i\in S\)). Note that this differs from \(x_i\), which
corresponds to the density of individuals of type \(i\) in the infinite
population for the RD equation. The stochastic process is
defined by the following steps:

\begin{enumerate}
    \item For a given population vector \(v_i\) the fitness vector \(f=f(v)\) is
        calculated.
    \item An individual from the population is randomly selected for copying.
        The probability of selecting an individual of type \(i\) is proportional
        to the fitness and given by:
        \begin{equation}\label{eqn:moran_pcopy}
        P_\text{copy}(i) = \frac{f_i(v)}{\sum_{i}f_i(v)}
        \end{equation}
    \item An individual from the population is randomly selected for removal.
        The probability of selection is uniform, that is:
        \begin{equation}\label{eqn:moran_premoval}
        P_\text{removal}(i) = \frac{1}{M}
        \end{equation}
    \item The individual selected for removal is removed and a new individual is
        introduced that is of the same type as the individual selected for
        copying.
\end{enumerate}

We repeat these steps until the population is uniform: that is, until all
individuals are of the same type. We consider the probability of any given type taking over, that is, the
fixation probability. The Moran process captures the role of stochastic
drift in finite populations: an individually fitter strategy is more likely,
but not guaranteed, to take over. Computational aspects of evolutionary dynamics in finite populations are
discussed in~\cite{hindersin2019computation}.

\section{The General strategic jokeying model}
\label{sec:general_model}

We introduce the jockeying model that will serve as the running example for
the rest of the paper: a representative case of a strategic queueing problem
for which no closed-form fitness expression is available, and which
therefore motivates the DEPU framework.

To motivate the model, consider a student arriving at a multi-storey
university library and looking for a free seat to work. The library has
several floors, each with a different number of seats, and seats free up at
different rates throughout the day. The student must decide which floor to
visit first, how long to wait there before giving up and moving to another
floor, and so on. The student's experience, measured in total time to find a
seat, depends both on their own search strategy and on the strategies used
by every other student. If most students start on the top floor and work
their way down, then a student who starts at the bottom might be likely to find a
seat sooner. The model in this section formalises this kind of situation as
a queueing system, and the rest of the paper is concerned with finding efficient
approaches to identifying
which search strategies emerge when many such individuals interact.

Consider a queueing system with \(N\) parallel service centres, indexed by
\(i \in \{1, 2, \dots, N\}\), each with service distribution \(G_i\) and
number of servers \(c_i\). Individuals arrive according to a Poisson process
with rate \(\Lambda\). Each individual has a strategy \(s=(\pi, j) \in S'
\subseteq S = [N] ^ n \times \left(\mathbb{R}^{+}\right)^n\) denoting a
sequence \(\pi\) of nodes from which to attempt to get service, and a
jockeying limit \(j_k\) for the \(k\)th visit in \(\pi\). In general \(n\)
can be any positive integer, which implies that individuals can return to a
node, and that they might wait a different amount of time for each visit to a
node; the jockeying limit is therefore indexed by position in the sequence
rather than by node label.

As an example of a strategy space, consider a system with \(N=2\) nodes and
restrict the strategy space to two types of behaviour:

\begin{equation}
    S'=\left\{\left((1, 2), (3, 9)\right), \left((2, 1), (6, 6)\right)\right\}
    \label{equ:small_example}
\end{equation}

The sum of the jockeying times for both types of individuals is 12:

\begin{itemize}
    \item the first strategy checks the first node and waits for 3 time
        units before waiting for at most 9 time units at the second;
    \item the second strategy checks the second node and waits for 6 time
        units before waiting for at most 6 time units at the first.
\end{itemize}

\begin{figure}[htbp!]
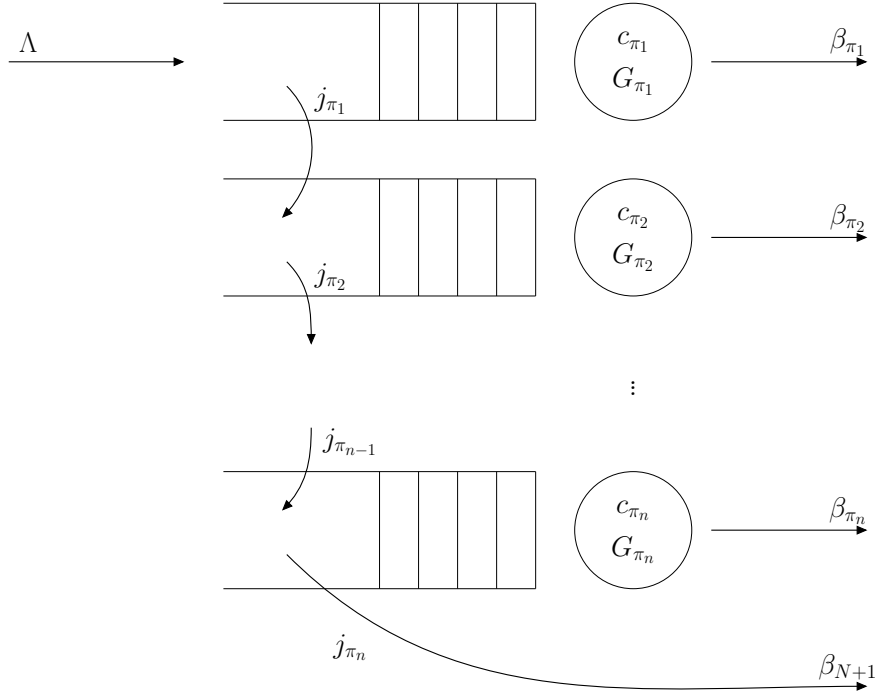

    \begin{center}
        \includestandalone[width=.7\textwidth]{jockeying_diagram}
        \caption{The multi-node jockeying queueing system.}
        \label{fig:jockeying}
    \end{center}
\end{figure}

The possibilities for strategy spaces are endless, so in practice we consider a
sensible subset \(S'\) that requires \(\pi\) to be a
permutation of \([N]\), so that each node is visited once. This model is
shown in Figure~\ref{fig:jockeying}.

An individual with strategy \((\pi, j)\) first visits node \(\pi_1\) and
waits a maximum time \(j_1\) for service. If they do not begin service
before \(j_1\) time units, they move to node \(\pi_2\), and so on.
If they receive service at node \(\pi_k\), then their
cost is the amount of time spent in the system, either including the time
spent in service or not.
After obtaining
service at node \(\pi_k\) they incur an exit cost of \(\beta_{\pi_k}\). If,
after visiting all the nodes, they are yet to receive service, they are lost
to the system and incur a cost of \(\beta_{N + 1}\) plus the amount of time
spent in the system.

For a particular individual progressing through the system, define the following
random variables:

\begin{enumerate}
    \item \(W_{\pi_k}\), the amount of time spent waiting at node \(\pi_k\); 
    \item \(\tau_{\pi_k} = 
          \begin{cases}
              \epsilon g_{\pi_k} + \beta_{\pi_k} & \text{if they received
              service at node }\pi_k,\\
              0 & \text{otherwise;}\\
          \end{cases}\)\\
          where \(g_{\pi_k} \sim G_{\pi_k}\) is the service time received at node
          \(\pi_k\), and \(\epsilon \in \{0, 1\}\) indicates whether a customer
          values their service time or not; 
    \item \(L = \begin{cases}0 & \text{if they received service,} \\ 1 & \text{if they were lost to the system.}\end{cases}\)
\end{enumerate}

Note that for a given strategy \((\pi, j)\) we have \(0 \leq W_{\pi_k} \leq
j_k\), \(\tau_{\pi_k} = 0\) for all service centres where the individual
did not receive service, and \(L=1\) if and only if \(\tau_{\pi_k} = 0\) for all
\(k\). Note also that \(\mathbb{E}(L)\) is the probability that the given
individual is lost to the system.

The cost to an individual is given by:

\begin{equation}\label{equ:cost}
    C = L\beta_{N + 1} + \sum_{k=1}^n \left( W_{\pi_k} + \tau_{\pi_k} \right)
\end{equation}

For the queueing model under consideration the `population' is transient: that
is, customers enter and leave the system as part of their interaction with it.
We need to define precisely what a population of strategies is in this context.
Consider a population of strategies that exists outside the queueing system,
large relative to the size of any transient population interacting with the
queueing system, and that feeds customers into the queueing system. Let
\(x\) represent the proportion of individuals in this larger
population that follows each strategy, thus \(x_s\) denotes the proportion of
individuals using strategy \(s\). We assume that:

\begin{equation}
    \sum_{s\in S'}x_s = 1
\end{equation}

Assume that this population randomly sends individuals to the queueing system.
Due to Poisson thinning of the arrival process the effective arrival rate
for customers following strategy \(s \in S\) at the queueing system is then
\(\Lambda_s = x_s \Lambda\); we assume throughout that arrivals are Poisson,
so that this decomposition holds.

To study emergence of behaviour in an evolutionary setting,
instead of cost, where a lower cost is beneficial, we consider fitness. The
fitness of a strategy \(s\in S'\), denoted by \(f_s\), is an order inverting
mapping to the positive numbers of the expectation
of~(\ref{equ:cost}):

\begin{equation}
    f_s = f_{\pi, j} = e ^ {- \kappa \mathbb{E}(C)} =
            e ^{- \kappa \left(\displaystyle{\mathbb{E}(L) \beta_{N + 1} + \sum_{k=1}^n
            \left( \mathbb{E}(W_{\pi_k}) + \mathbb{E}(\tau_{\pi_k})\right)}\right)} \label{equ:population_fitness}
\end{equation}

Here \(\kappa\) is referred to as the selection
intensity~\cite{hindersin2019computation}. A value of \(\kappa \ll 1\) implies
that congestion has a low effect on emergent behaviour, which becomes random.

\subsection{Equivalence to model with travel times}
\label{sec:model_with_travel_times}

Consider a modification of the system in which individuals experience
travel time from:
\begin{itemize}
    \item arrival to node;
    \item node to node;
    \item abandonment to exit.
\end{itemize}

This is given by a matrix \(T\in\mathbb{R}^{(N + 1) \times (N + 1)}\) where:

\begin{itemize}
    \item \(T_{N + 1, j}\) gives the travel time from arriving to node \(j\);
    \item \(T_{ij}\) gives the travel time from node \(i\) to node \(j\);
    \item \(T_{i, N + 1}\) gives the travel time from node \(i\) to exiting the
        system.
\end{itemize}

This can be modelled using the general jockeying model described in
Section~\ref{sec:general_model} with the addition of a holding node 
as the \((N+1)\)th node with \(c_{N+1} = 0\), together with a constraint on the
strategy space \(\bar S \subseteq S\):

\begin{equation}
    \bar S = \left\{(\pi, j) \in S \;\left|\; 
    \begin{array}{l}
        \pi \in [N + 1]^{2N + 1}, \text{and } \pi_i = N + 1 \text{ for \(i\) odd}\\
        j \in \mathbb{R}^{+}, \text{and } j_i = T_{\pi_{i-1}, \pi_{i+1}} \text{ for \(i\) odd}
    \end{array}\right.\right\}
\end{equation}

The odd-indexed elements of \(j\) correspond to the travel times.

Before travelling to a given node in their strategy, the individual first
travels to the holding node, where they will not be served (since
\(c_{N + 1} = 0\)) and so will leave after the corresponding jockeying limit.
Thus, the individual queues at the holding node for a given travel time and
then jockeys to the next node.

For example, consider the system with strategy space given by
Equation~(\ref{equ:small_example}). Suppose the following travel times need to
be considered:

\begin{itemize}
    \item Travel from arrival to the first node takes 7 time units.
    \item Travel from arrival to the second node takes 6 time units.
    \item Travel from the first node to the second takes 15 time units and from
        the second to the first takes 4 time units.
    \item Exit from the first node takes 12 time units and exit from the second
        takes 5 time units.
\end{itemize}

The travel time matrix \(T\) is:

\begin{equation}
    T = \begin{pmatrix}
        0& 15& 12\\
        4& 0& 5\\
        7& 6& 0
    \end{pmatrix}
    \label{equ:small_example_travel_time_matrix}
\end{equation}

The strategy space is then given by:

\begin{equation}
    S'=\left\{\left((3, 1, 3, 2, 3), (7, 3, 15, 9, 5)\right),
              \left((3, 2, 3, 1, 3), (6, 6, 4, 6, 12)\right)\right\}
    \label{equ:small_example_with_travel_times}
\end{equation}

The model just described is rich enough to capture a range of practical
situations. The library example of Section~\ref{sec:general_model} is one;
others include patients choosing between hospitals with different waiting
times and travel costs, individuals choosing evacuation strategies, 
or commuters choosing between rival lanes at a toll
plaza. In each case the strategy space and travel time matrix differ, but
the underlying question is the same: given a population of strategic
individuals, each searching for service from a collection of parallel
queues, which search strategies emerge?

More broadly, any complex queueing system in which an individual's payoff
depends on the strategies of others belongs to the same class of problem.
The jockeying model is one instance. In each case the strategy space,
cost function, and system dynamics differ, but the underlying question is
the same: which strategies emerge when many self-interested individuals
interact in a stochastic queueing environment? Whenever a DES of the system can be constructed, DEPU provides a
general answer.

In Section~\ref{sec:baseline} we apply classical evolutionary dynamics to
this model via DES; in Section~\ref{sec:depu} we introduce DEPU and
demonstrate it on the same model at substantially lower computational cost.

\section{Applying classical evolutionary dynamics via DES}
\label{sec:baseline}

We first apply the classical RD and Moran process methods to the jockeying
model using DES to estimate fitnesses. This baseline establishes that these
methods can in principle solve the problem, and the computational
inefficiency we observe motivates the DEPU framework presented in
Section~\ref{sec:depu}.

We begin with how to evaluate the fitness function without a closed-form
formula, using simulation, described in Section~\ref{sec:des}. We then
consider its behaviour under the RD model in
Section~\ref{sec:apply_replicator_dynamics}, and under the Moran process in
Section~\ref{sec:apply_moran_process}.

\subsection{Evaluating fitness with discrete event simulation}\label{sec:des}

For the generalised model given in Section~\ref{sec:general_model}, closed
formulae for the fitnesses given a particular population vector are unknown.
Discrete event simulation (DES) is a common methodology for approximating key
performance measures in complex systems, and is especially popular when
considering queueing systems with complex behaviours. It is therefore natural
to use DES to find the fitnesses \(f_s(x)\) for a particular
population vector.

\begin{figure}[!hbtp]
    \centering
    \includegraphics[width=\textwidth]{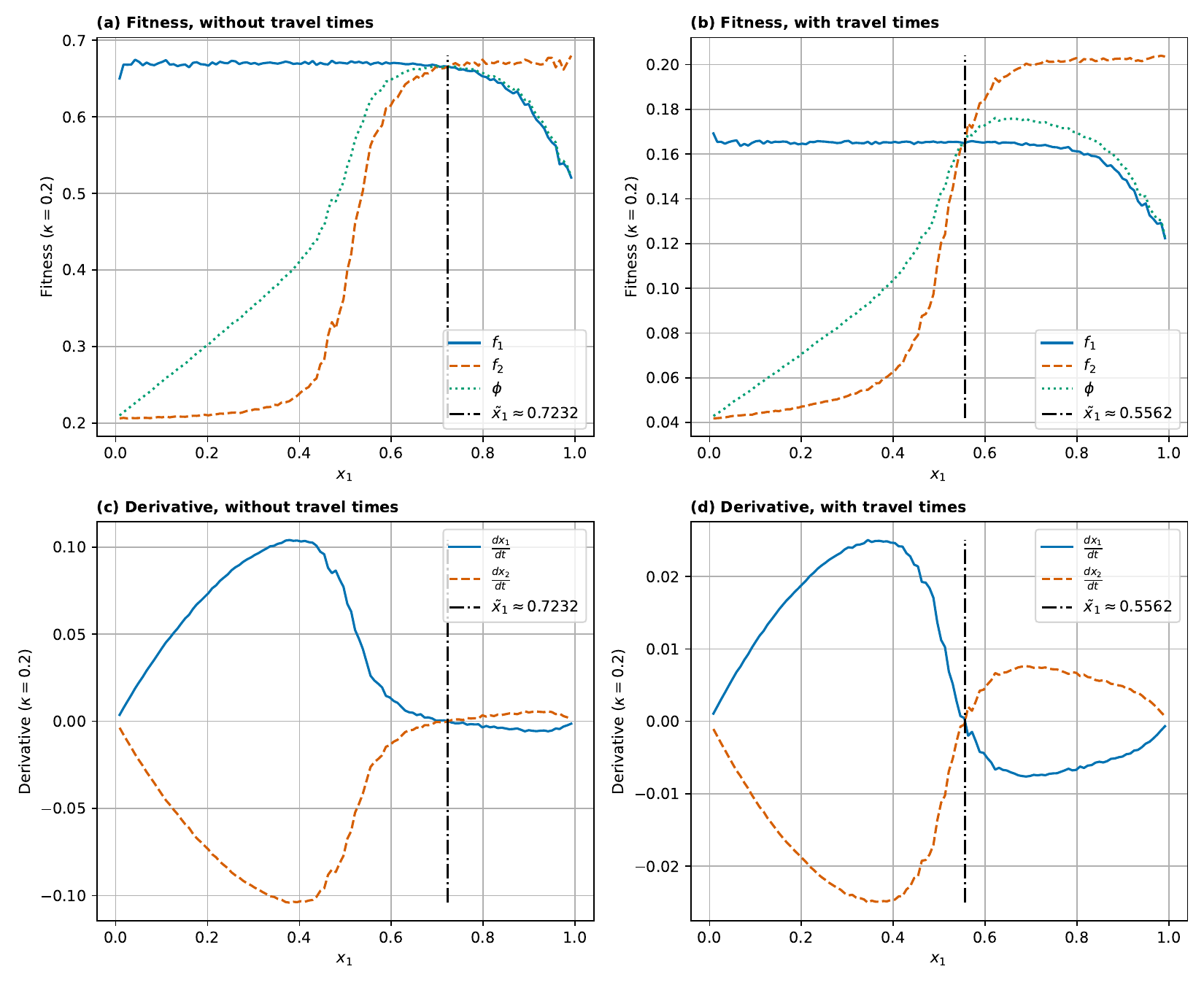}
    \caption{Fitness and replicator dynamics derivative for \(\Lambda=10,
    c_1=20, c_2=10, \mu_1=0.5, \mu_2=0.5, \beta_1=0, \beta_2=0, \beta_3=10\).
    Each fitness value is estimated by averaging 10 DES trials of
    \(12{,}000\) customers, with the first 100 customers of each trial
    discarded as warmup; the population proportion \(x_1\) is sampled on a
    grid of \(120\) points.
    (a) and~(b) show the fitness of each strategy as a function of
    the population proportion \(x_1\); the stable population corresponds to
    the intersection of \(f_1\) and \(f_2\). (c) and~(d) show the
    corresponding replicator dynamics derivative; the derivative is zero at
    the same mixed equilibrium. In each case (a) and~(c) show the
    scenario without travel times, and (b) and~(d) show the scenario
    with travel times (Section~\ref{sec:model_with_travel_times}) given by the
    travel time matrix in
    Equation~\eqref{equ:small_example_travel_time_matrix}.}
    \label{fig:two_node_example_for_given_population_densities}
\end{figure}

To demonstrate finding \(f_s(x)\) using DES, consider a system
with \(N=2\), and the strategy space given in equation~\eqref{equ:small_example}.
Figure~\ref{fig:two_node_example_for_given_population_densities}(a)
shows the overall fitness for a given population \(x=(x_1, x_2)=(x_1,
1-x_1)\) evaluated using DES.
Figure~\ref{fig:two_node_example_for_given_population_densities}(b)
is a slight modification, described in
Section~\ref{sec:model_with_travel_times}, where travel times to and between
service nodes are included according to the travel time matrix in
Equation~\eqref{equ:small_example_travel_time_matrix}.

For completeness, and as preparation for Section~\ref{sec:depu}, we outline
implementation details of the DES in
Section~\ref{sec:simulation_implementation}.

\subsection{Implementation details}\label{sec:simulation_implementation}
\begin{figure}[!htbp]
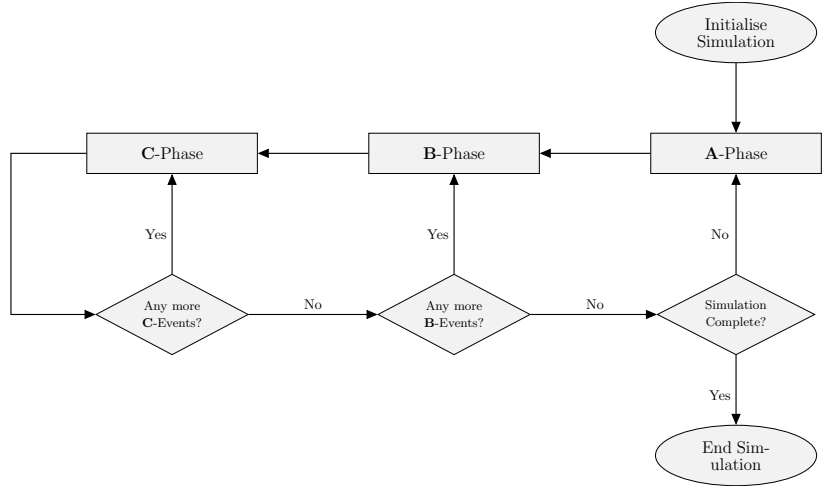

    \begin{center}
        \includestandalone[width=0.65\textwidth]{event_scheduling_diagram}
    \end{center}
    \caption{Diagram of the three phase event scheduling approach used in Ciw.}
    \label{fig:event_scheduling}
\end{figure}

A standard implementation of DES uses the three-phase
approach~\cite{pidd98, robinson14}. Here events occur throughout the run of
the simulation and are categorised into \textbf{B}-events, or scheduled
events, and \textbf{C}-events, or unscheduled events. In this work we use the
\texttt{Ciw} library~\cite{ciw3.2.4, palmer2019}, which implements this as
shown in Figure~\ref{fig:event_scheduling}, in a slight variation to the
standard whereby triggered \textbf{C}-events are carried out as they are
triggered, rather than waiting for all \textbf{B}-events to complete. For the
queueing system described in Section~\ref{sec:general_model}, the
\textbf{B}-events include customers arriving (scheduled a sampled time
interval after the previous arrival), customers finishing service (scheduled
a sampled time interval after the customer began service), and customers
deciding to jockey to another queue or leave the system (scheduled a sampled
time interval after the customer entered that node). The \textbf{C}-events
include a customer beginning service (triggered by another customer leaving
service). Note that some events can change or cancel future scheduled events;
for example, if a customer begins service, then their jockeying event is
cancelled.

\subsection{Replicator Dynamics}\label{sec:apply_replicator_dynamics}
We now apply the RD model of evolutionary behaviour described
in Section~\ref{sec:replicator_dynamics} to our jockeying queueing model.
This involves finding a population \(x\) such that the derivative
given by Equation~\eqref{eqn:replicator_dynamics_equation} is equal to zero.

The derivative itself can be found by evaluating the right hand side of
Equation~\eqref{eqn:replicator_dynamics_equation}, by evaluating the fitnesses
themselves using DES. Figure~\ref{fig:two_node_example_for_given_population_densities}
shows both the fitness values and the corresponding derivatives. We find three
values of \(x\) at which the derivative is zero: two corresponding
to pure equilibria where the population consists of only one strategy; and one
mixed equilibrium, corresponding to a specific proportion of the population
following one strategy and the rest following the other.

In the present two-strategy example the population is described by the
single proportion \(x_1\), so the equilibria can be located simply by
evaluating the derivative on a grid, as in
Figure~\ref{fig:two_node_example_for_given_population_densities}. For a
larger strategy space this becomes impractical: gridding all of \(x\) is
inefficient. There exist a number of algorithms that can
solve differential equations such as
Equation~\eqref{eqn:replicator_dynamics_equation} numerically; for example
Euler's method, Runge-Kutta methods, and multi-step methods~\cite{gautschi11}.
For the purposes of this work Euler's method is sufficient. It is
an iterative algorithm that defines the value of \(x\) at some time step
\(x(t + \Delta t)\) as a function of the value of \(x\) at the previous time
step \(x(t)\) using the following:
\begin{equation}
    x_i(t +  \Delta t) = x_i(t) + \Delta t \frac{dx_i(t)}{dt}
    \label{eqn:euler_method_time_update}
\end{equation}

where \(\Delta t\) is an arbitrary time increment.

\begin{figure}[!hbtp]
    \centering
    \includegraphics[width=\textwidth]{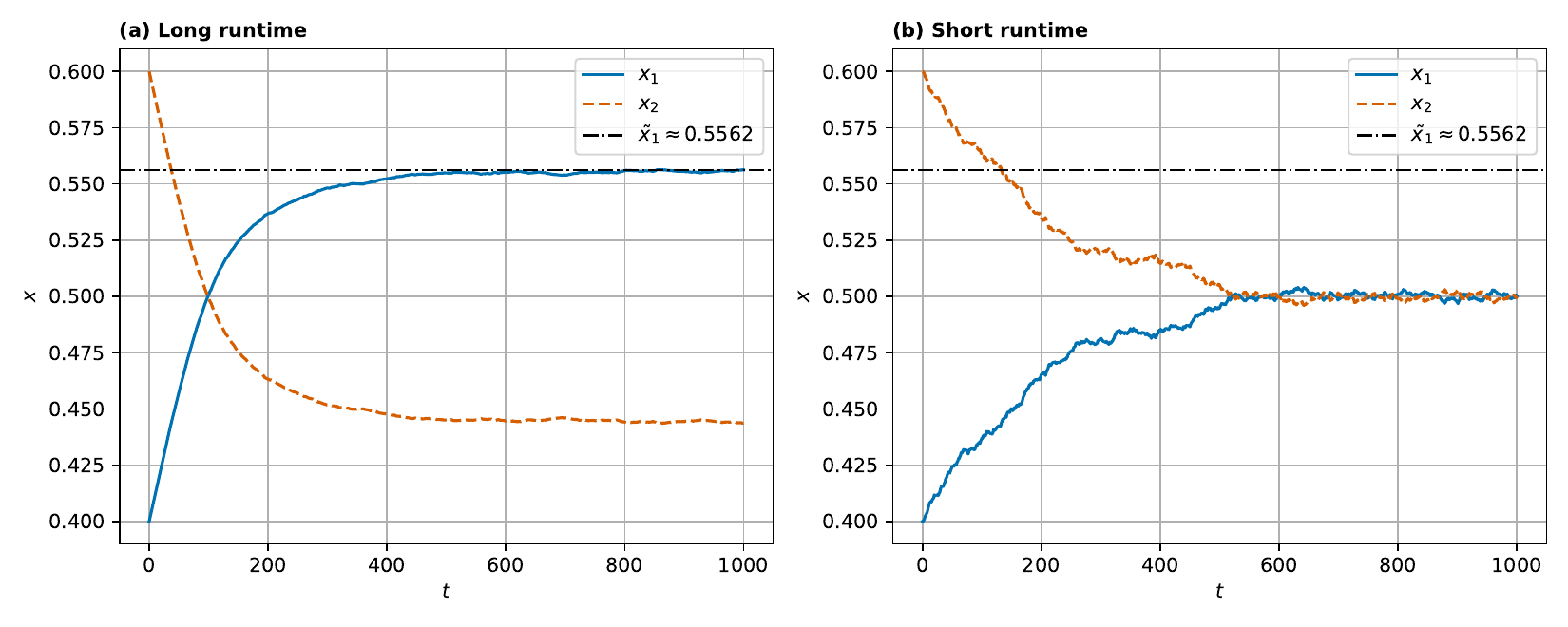}
    \caption{The emergent behaviour for the example considered in
    Figure~\ref{fig:two_node_example_for_given_population_densities}(b).
    (a): \(f\) is evaluated with a long DES run of \(20{,}000\) customers;
    the population stabilises at \(\tilde x\approx(0.5557, 0.4443)\).
    (b): \(f\) is evaluated with a far shorter run of \(200\) customers;
    the reduced accuracy causes the method to converge to a different,
    incorrect value, \(\tilde x\approx(0.5002, 0.4998)\). In both panels the
    dash-dotted black line marks the mixed equilibrium
    \(\tilde x_1\approx 0.5562\) located from the same fitness data as
    Figure~\ref{fig:two_node_example_for_given_population_densities}; the long
    run converges to it, whereas the short run does not.}
    \label{fig:apply_replicator_dynamics}
\end{figure}

Figure~\ref{fig:apply_replicator_dynamics}(a) shows the population
over time evaluated using Euler's method, for the example with travel times
described in Section~\ref{sec:model_with_travel_times}. We see that the
method converges to a value of \(\tilde x\approx(0.5557, 0.4443)\),
matching the mixed equilibrium \(\tilde x_1\approx 0.5562\) located from the
fitness data in
Figure~\ref{fig:two_node_example_for_given_population_densities}(d) and drawn
as the dash-dotted line in Figure~\ref{fig:apply_replicator_dynamics}(a).
A long run of the DES was used to obtain a good approximation of the
fitnesses, and therefore of the derivative \(\frac{dx_i(t)}{dt}\) from
Equation~\eqref{eqn:replicator_dynamics_equation} at each time step. This is
computationally expensive; however, a long simulation time is required to
obtain a good approximation of the fitness, since this method is sensitive to
the accuracy of the approximations. For example, when reducing the simulation
run from \(20{,}000\) customers to \(200\) customers, and thus reducing the
accuracy of the fitnesses, the method converges on a different, incorrect
value. In Section~\ref{sec:depu} we present an approach that accurately
converges to an equilibrium behaviour at a much lower computational cost.
 
\subsection{Moran process}\label{sec:apply_moran_process}
We now apply the Moran process model of evolutionary behaviour described in
Section~\ref{sec:moran_process} to our jockeying queueing model. This involves
tracking the population \(v_i\) over a number of iterations, typically until
one strategy takes over the population. This behaviour is particularly
sensitive to the total population size \(M = \sum_{s\in S} v_s\).

\begin{figure}[!hbtp]
    \centering
    \includegraphics[width=\textwidth]{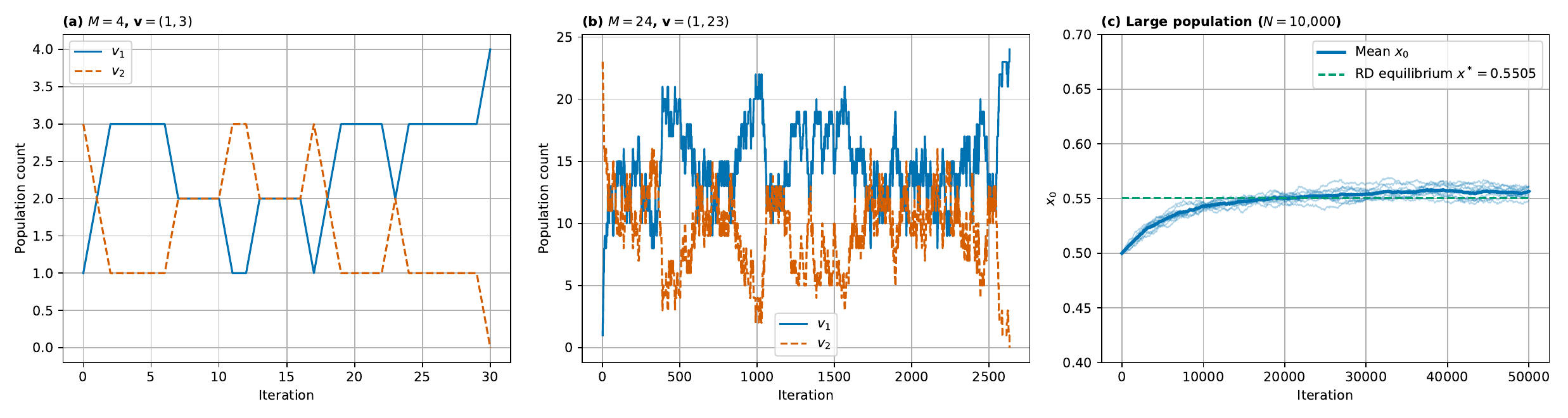}
    \caption{Examples of runs of the Moran process and large-population
    convergence. At each Moran iteration the two strategy fitnesses are
    estimated by a single DES trial of \(10{,}000\) customers with the
    first 200 customers discarded as warmup. (a) and~(b) show individual
    trajectories from initial populations \((v_1,v_2)=(1,3)\) and
    \((1,23)\) respectively; in each case the process ends when one
    strategy takes over. (c) shows ten independent seeds of a large Moran
    process (\(M=10{,}000\), \(50{,}000\) iterations each, initial
    population \((5{,}000, 5{,}000)\)); the proportion
    \(x_1 = v_1 / M\) settles around the equilibrium of the RD. (c) plots
    \(x_1\) rather than the raw count \(v_1\), so the quantity is directly
    comparable with the RD equilibrium \(x^*\).}
    \label{fig:moran_process_examples}
\end{figure}

Figure~\ref{fig:moran_process_examples} shows Moran processes for different
population sizes and different starting populations. In each case, the Moran
process ends when one strategy takes over the population. The probability of
any given type taking over is referred to as the fixation probability, and
can be found by running many trials of the same Moran process and recording
which strategy took over the population each time. After 2000 trials, the
fixation probabilities for these two starting populations are given in
Table~\ref{tbl:moran_process_examples_fixation_probabilities}.

\begin{table}[!hbtp]
\centering
\caption{The fixation probabilities for the scenario of
Figure~\ref{fig:moran_process_examples}.}
\label{tbl:moran_process_examples_fixation_probabilities}
\begin{tabular}{rr}
\toprule
Initial population & Probability of first type taking over \\
\midrule
(1, 3) & 0.521500 \\
(1, 23) & 0.303500 \\
\bottomrule
\end{tabular}
\end{table}

No fixation is shown for the case of \(N=10,000\) as no fixation arises, in fact
this experimentally demonstrates the result
of~\cite{traulsen2005coevolutionary}, that the Moran process converges to the
RD equation for asymptotic values of \(M\), by running this
Moran process for a large value of \(M\). This is illustrated in
Figure~\ref{fig:moran_process_examples}(c), where a Moran process for a large
population converges to the same population distribution as in
Figure~\ref{fig:two_node_example_for_given_population_densities}(b).

Both baseline approaches converge to the expected equilibrium behaviour, but
at a cost: each requires a separate DES run for every population update. We
address this nested-loop inefficiency in the next section.

\section{Discrete event population updates}\label{sec:depu}
\begin{figure}
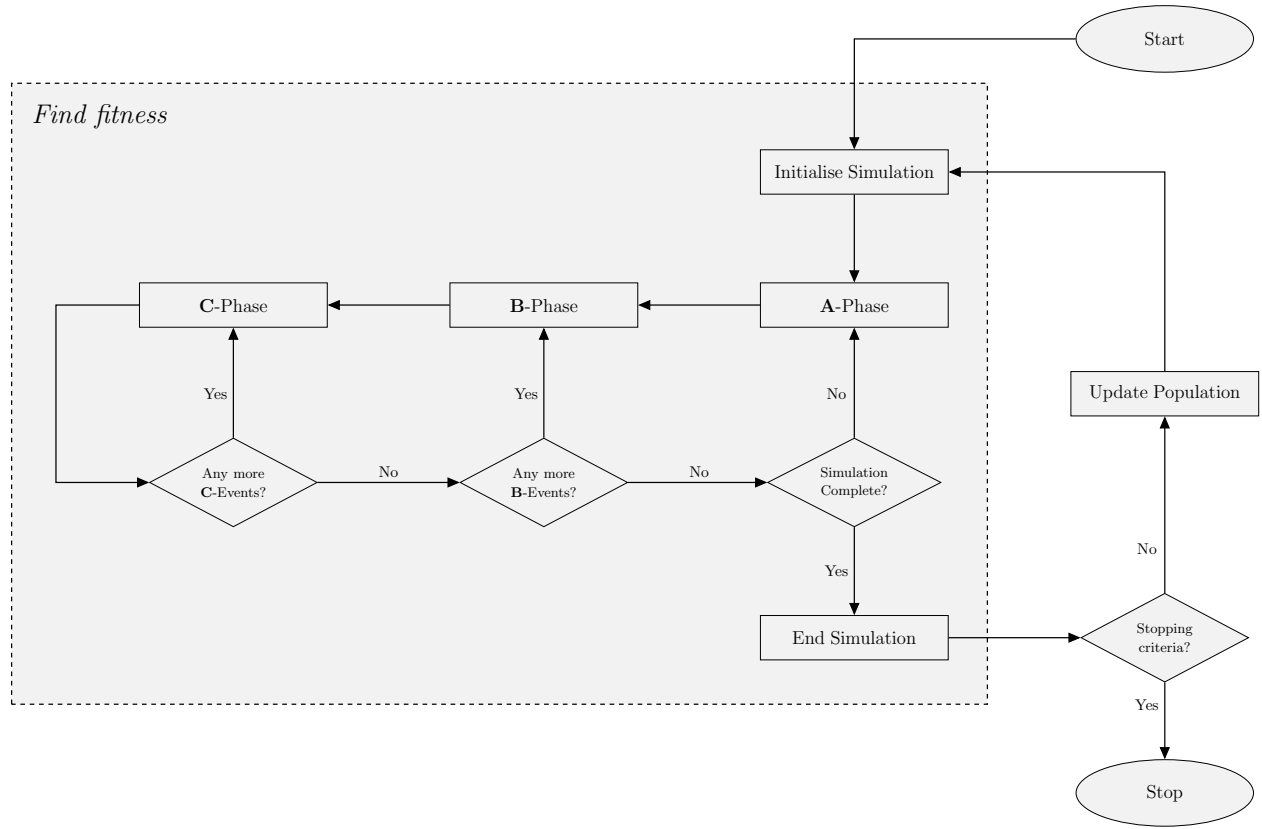
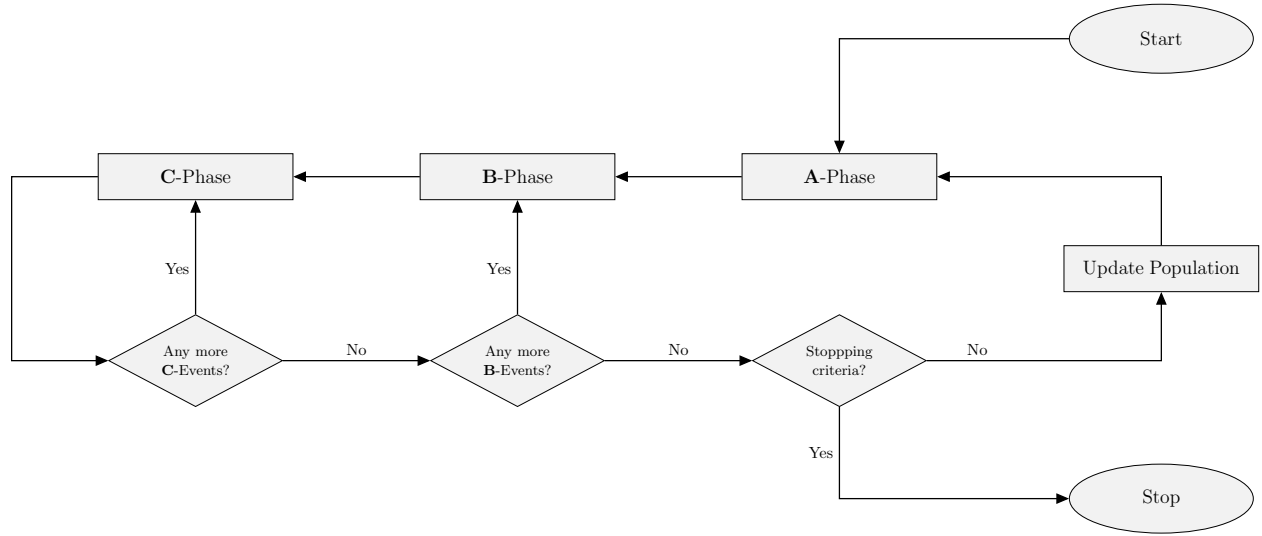

\begin{subfigure}{\textwidth}
\begin{center}
\includestandalone[width=\textwidth]{get_fitness_by_simulation}
\end{center}
\caption{Diagrammatic representation of population updates when using event scheduling to find fitnesses.}
\label{fig:update_population_event_scheduling}
\end{subfigure}
\begin{subfigure}{\textwidth}
\vspace{10mm}
\begin{center}
\includestandalone[width=\textwidth]{depu_diagram}
\end{center}
\caption{Diagrammatic representation of DEPU.}
\label{fig:depu}
\end{subfigure}
\caption{Diagrammatic representation comparing using DEPU and the using
DES to approximate fitnesses in an emergent population. Note that the DEPU
flow of Figure~\ref{fig:depu} is a re-wiring of
Figure~\ref{fig:update_population_event_scheduling}.}
\label{fig:depu_vs_des}
\end{figure}

The two methods used above for finding emergent behaviour, RD
(Section~\ref{sec:apply_replicator_dynamics}) and the Moran process
(Section~\ref{sec:apply_moran_process}), both follow the same iterative
process of evaluating and updating a population. When the evaluation is done
through DES, described in Section~\ref{sec:des}, this
iterative process can be represented diagrammatically as in
Figure~\ref{fig:update_population_event_scheduling}.

This nests an iterative process (the DES) within another iterative process
(the population updates). The inner iterative process, the DES, uses the law
of large numbers across iterations to refine the estimate of
\(f_s(x)\) for a given \(x\). In practice this is
calculated by generating many individuals across the strategies, each
receiving a cost according to Equation~\eqref{equ:cost} that is mapped to a
fitness by Equation~\eqref{equ:population_fitness}, and then taking an
average. The more individuals that are generated, the better the estimate of
the average fitness. The outer iterative process follows game-theoretic rules
to update the population \(x\) until convergence to some emergent
equilibrium.

This nested iterative process can be inefficient and slow: the inner DES is
re-run at every outer step, even though the population vector changes only
slightly between consecutive outer iterations.

DEPU re-wires the diagram in
Figure~\ref{fig:update_population_event_scheduling} to avoid the nested
iterative process. The diagrammatic representation for
DEPU is given in Figure~\ref{fig:depu}. Here, populations are updated each
time the clock of the simulation advances to the next relevant scheduled
\textbf{B}-event, in our case each time a customer leaves the system and
receives a cost. This means that both the fitnesses and population vectors
are updated simultaneously, allowing for a single long run of the simulation
model instead of numerous runs to approximate the fitness function.

\subsection{Estimating fitness with DEPU}
The update methods for the fitnesses and for the population vectors are
different. We receive new information about the fitnesses each time a
customer of strategy \(s\) leaves the system and incurs a cost \(C_s\), which
is a random variable. From this single data point one estimate of the fitness
would be \(f_s = e^{-\kappa C_s}\)
as in~\eqref{equ:population_fitness}. However this would also be a random
variable, and would differ from previous estimates of the fitness both
because of stochasticity, and because the population vector will have been
updated since the previous fitness update. The fitness update therefore needs
to both smooth out variability and learn new information based on the updated
population. This problem of estimating the expected value of a random
variable that changes over time is equivalent to a non-stationary multi-arm
bandit problem~\cite{sutton2018reinforcement}, and we therefore propose
updating the estimate for the fitness with a learning rate, or exponential
smoothing update, traditionally used for these types of problems.

Let \(\alpha\) denote a learning rate, that is, the relative importance of
the most recent update in comparison to previous updates. The exponential
smoothing update is given in
Equation~\eqref{eqn:exponential_smoothing_update}, applied each time an
individual incurs a cost, where \(C_s\) is the cost received by the current
individual of strategy \(s\).

\begin{equation}\label{eqn:exponential_smoothing_update}
    f_s \leftarrow e ^ {-\kappa \left(\alpha C_s + (1 - \alpha) \mathbb{E}(C_s)\right)}
\end{equation}

This is analogous to~\eqref{equ:population_fitness}, where
\(\mathbb{E}(C_s)\) corresponds to the current best known estimate of the
expected cost of strategy \(s\). In practice we do not need to keep track of
this, as the current estimate of the fitness can be mapped back to the cost
by taking the inverse function of~\eqref{equ:population_fitness}. This gives:

\begin{align}
    f_s &\leftarrow e ^ {-\kappa \left(\alpha C_s + (1 - \alpha) \frac{\ln(f_s)}{-\kappa}\right)}\\
    f_s     &\leftarrow e ^ {-\kappa \alpha C_s} f_s^{(1 - \alpha)}
\end{align}

As noted in Figure~\ref{fig:depu}, the population vector can be updated each
time the fitnesses are updated. The population vector updates can be carried
out in analogous ways to the RD
(Section~\ref{sec:replicator_dynamics}) or the Moran process
(Section~\ref{sec:moran_process}):

\begin{itemize}
    \item when the population vector update corresponds to using a numerical
        method to solve the RD equation, we call this Discrete
        Event Replicator Dynamics (DERD);
    \item when the population vector update corresponds to randomly removing
        or reproducing strategies in the population vector, analogously to the
        Moran process, we call this Discrete Event Moran Replacement (DEMR).
\end{itemize}

\subsection{DERD: Discrete Event Replicator Dynamics}\label{sec:derd}
In this method, the population vector \(x\) is updated according
to some numerical method for solving the RD equation. We use
the same rule as Euler's method~\eqref{eqn:euler_method_time_update}, defining
the derivative using the RD
equation~\eqref{eqn:replicator_dynamics_equation}, which gives:

\begin{align}\label{eqn:fitness_to_prob}
  x_s &\leftarrow x_s + \Delta t \frac{dx_s}{dt}\\
  x_s &\leftarrow x_s + \Delta t x_s\left(f_s - \phi\right)
\end{align}

Figure~\ref{fig:derd_demr_examples}(a) shows the emergent behaviour of the
model instance presented in Section~\ref{sec:model_with_travel_times} when
using DERD. The population converges efficiently to the equilibrium value
obtained using numerical integration, shown with dotted lines. Note that, in
contrast to
Figure~\ref{fig:two_node_example_for_given_population_densities}(b),
the x-axis represents \emph{simulation time} of one long run of the
simulation, and not the number of iterations of Euler's method.

At the beginning of the DERD run, the current estimates for the fitnesses are
poor, and each new observation may change the current estimate greatly: this
means that at the beginning of the run the populations may change
erratically. By the end of the run we have good estimates of the fitnesses;
however, the new observations are still stochastic random variables, so
depending on the learning rate the estimates of the fitnesses may continue to
fluctuate, and therefore so do the populations. A sensible choice of
hyperparameters can alleviate this; further discussion of the effect of the
hyperparameters is given in Section~\ref{sec:hyperparameters}.

\subsection{DEMR: Discrete Event Moran Replacement}\label{sec:demr}
In this method, the population vector \(v\) is a vector of integer counts
of fixed total \(M = \sum_{s \in S} v_s\). Each time the fitnesses are
updated during the run of the DES (that is, each time a customer leaves
the system and incurs a cost), one Moran replacement step is applied to
\(v\). A strategy is selected for copying with probability
\(P_\text{copy}\) given by Equation~\eqref{eqn:moran_pcopy}, evaluated
using the current fitness estimates. A second strategy is selected for
removal uniformly at random as in Equation~\eqref{eqn:moran_premoval}.
The selected count is decremented by one and the copied count is
incremented by one, leaving \(M\) unchanged.

The corresponding strategy proportions \(x_s = v_s / M\) are then used by
the router in the DES until the next cost is
incurred, at which point a further replacement step is applied. In contrast to the
classical Moran process of Section~\ref{sec:moran_process}, the
fitnesses used here are the running DEPU estimates rather than fresh
DES estimates, so DEMR avoids the nested inner simulation entirely.

\begin{figure}[!hbtp]
  \centering
  \includegraphics[width=\textwidth]{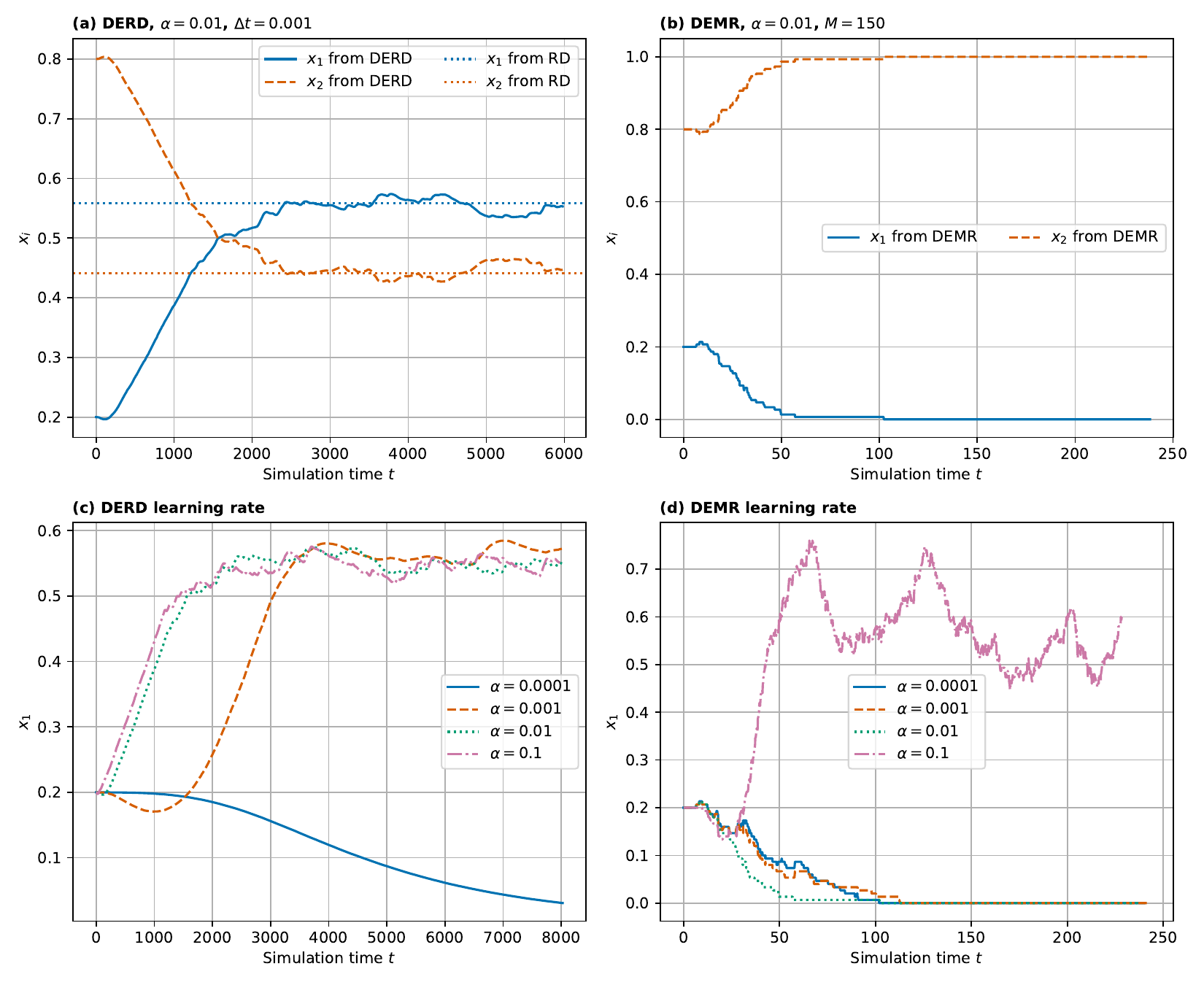}
  \caption{Emergent behaviour of the model from
    Section~\ref{sec:model_with_travel_times} under DERD and DEMR, and the
    effect of the learning rate \(\alpha\). (a) and~(b) show example
    trajectories for DERD (\(\alpha=0.01\), \(\Delta t=0.001\)) and DEMR
    (\(\alpha=0.01\), \(M=150\)) respectively; both converge to the same
    equilibrium, shown with dotted lines in~(a). (c) and~(d) show the
    effect of varying \(\alpha\) for DERD and DEMR respectively.}
  \label{fig:derd_demr_examples}
\end{figure}

Figure~\ref{fig:derd_demr_examples}(b) shows the emergent behaviour of the
same model instance when using DEMR.

\subsection{Effect of algorithm parameters}\label{sec:hyperparameters}
Both DERD and DEMR contain hyperparameters that can greatly affect their
behaviour:

\begin{itemize}
  \item both algorithms use a learning rate \(\alpha\);
  \item DERD uses a time step increment \(\Delta t\);
  \item DEMR uses a population size \(M\).
\end{itemize}

Figure~\ref{fig:derd_demr_examples}(c) shows the effect of the learning rate
\(\alpha\) when using DERD on the scenario described in
Section~\ref{sec:model_with_travel_times}, plotting the population of the
first strategy \(x_1\) over time for different values of \(\alpha\). For
DERD, larger learning rates help the population find the stable point more
quickly, albeit with diminishing returns. The horizontal axis is simulation
time \(t\), in the time units of the underlying DES; with arrival rate
\(\Lambda=10\), the \(80{,}000\) customers used here correspond to
\(t \approx 8{,}000\), which is the range shown in panel~(c). The smallest
learning rate shown, \(\alpha=0.0001\), has not found a stable population
within this window; it is not clear from the figure which stable population
this learning rate would converge to, the mixed or pure equilibrium. For those runs where a stable population has been reached, the
populations fluctuate stochastically around the equilibrium; the larger the
learning rate, the less erratic these fluctuations, since a larger learning
rate smooths out the stochasticity. Figure~\ref{fig:derd_demr_examples}(d) repeats this experiment with DEMR.
For DEMR, small learning rates (\(\alpha \leq 0.01\)) consistently drive
the population to the pure equilibrium \(x_1 = 0\), with convergence speed
largely insensitive to the precise value of \(\alpha\). A large learning
rate (\(\alpha = 0.1\)) produces erratic dynamics that do not settle within
the simulation window. Notably, DEMR appears to select the pure rather than
the mixed equilibrium found by DERD. A plausible explanation is that DEMR,
like the classical Moran process, has absorbing states at the pure
populations, whereas DERD does not: once stochastic fluctuations drive a
strategy count to zero it cannot recover, so a pure population is eventually
reached and fixed. Making this argument precise is a natural open problem.

\begin{figure}
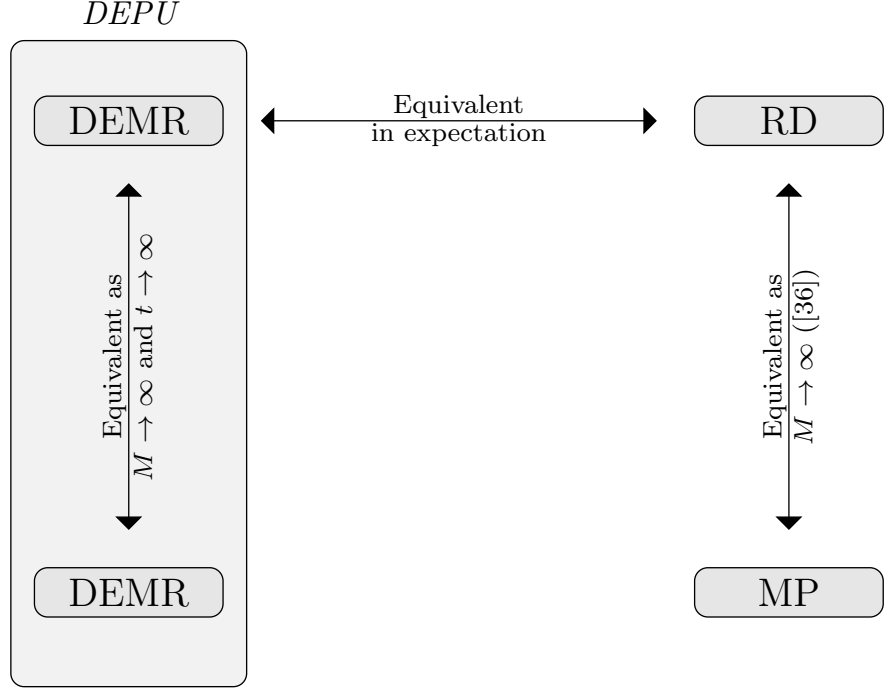

\centering
\includestandalone[width=0.7\textwidth]{compare_depu_rd_mp}
\caption{Diagram showing the relationships between DEPU, numerical solution of the replicator dynamics equation, and the Moran process.}
\label{fig:compare_depu_to_rd_and_mp}
\end{figure}

Figure~\ref{fig:compare_depu_to_rd_and_mp} shows how the four models
considered, that is, numerical solution of the RD equation, the Moran
process, DERD and DEMR, relate to one another. DERD recovers the RD
equation in the appropriate limit, just as the Moran process recovers the
RD equation as \(M \to \infty\); whether there are circumstances under
which DEMR and the classical Moran process coincide is a natural open
question.

\subsection{Application to Naor's observable queue}
\label{sec:recover_known_models}

\begin{figure}[!htbp]
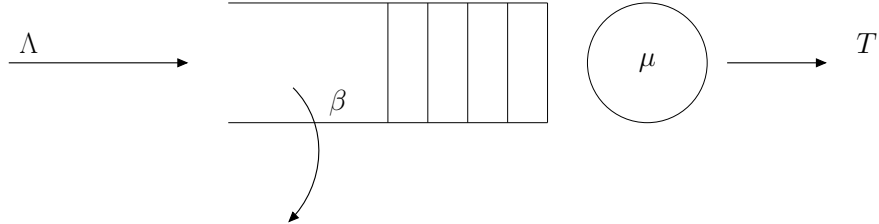

\begin{center}
\includestandalone[width=.7\textwidth]{naor_diagram}
    \caption{The queueing model of~\cite{naor1969regulation}. Individuals join or
    baulk based on a threshold strategy \(\tilde n\).}
\label{fig:naor}
\end{center}
\end{figure}

As a further example of DEPU in use, we apply it to a classic model in
behavioural queueing theory for which the equilibrium is already known
analytically, and verify that DEPU recovers it. We consider the queueing
model of~\cite{naor1969regulation}, depicted in
Figure~\ref{fig:naor}. Individuals arrive at a rate \(\Lambda\), observe the
queue size \(n\) of an \(M/M/1\) queue (a single-server queue with Markovian
inter-arrival and service rates), and then choose to join if and only if
\(n \leq \tilde n\). The value \(\tilde n\) is referred to as a threshold
strategy.

Individuals receive a cost based on the following:

\begin{itemize}
    \item A value of \(\beta\) if they choose to not join the queue.
    \item Their total time in the system if they join the queue.
\end{itemize}

One of the results proved in~\cite{naor1969regulation} is that the equilibrium
threshold strategy is given by:

\begin{equation}
    \tilde n = \lfloor \beta \mu \rfloor
    \label{equ:naor_equilibrium_threshold}
\end{equation}

\begin{figure}[!htbp]
    \begin{center}
        \includegraphics[width=.8\textwidth]{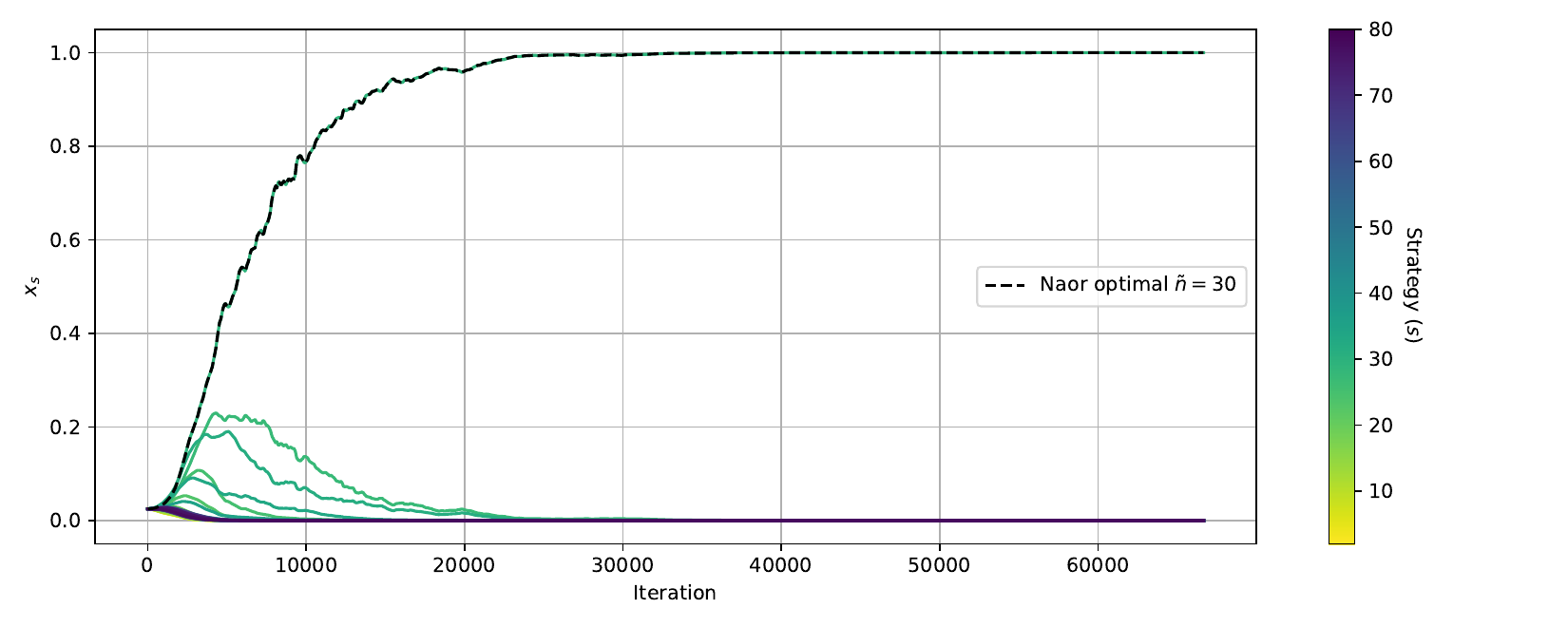}
    \end{center}
    \caption{Discrete Event Replicator Dynamics on Naor's observable M/M/1
    queue with \(\Lambda=30, \mu=20, \beta=1.5\). The strategy space is the
    set of even thresholds \(\{2, 4, \dots, 80\}\); the population proportion
    \(x_s\) of each threshold strategy is plotted over time, coloured by
    strategy value. Equation~\eqref{equ:naor_equilibrium_threshold} gives the
    Naor optimal threshold \(\tilde n = \lfloor \beta\mu \rfloor = \lfloor
    1.5 \cdot 20\rfloor = 30\); the trajectory for this strategy is overlaid
    as a dashed black line and converges to \(x_{30} = 1\), recovering the
    analytical result.}
    \label{fig:derd_applied_to_naor}
\end{figure}

Using Discrete Event Replicator Dynamics (Section~\ref{sec:derd}) this result
can be recovered. This is shown in Figure~\ref{fig:derd_applied_to_naor},
where the emergent behaviour corresponds
to~\eqref{equ:naor_equilibrium_threshold}.

The model of~\cite{naor1969regulation} is referred to as an ``observable''
model~\cite{shone2013comparisons}, in that individuals know the state of the
system before deciding whether to join or not. This is not the case with the
jockeying model of Section~\ref{sec:general_model}: individuals know how long
they have spent in each queue, but they do not know the state of the next
queue before deciding to jockey. Another example of an ``unobservable'' model
is considered in~\cite{bell1983individual}, where individuals do not know the
state of the system and use a probability strategy where they choose to join
an \(M/M/1\) queue with some probability. DEPU cannot be applied here without a specific modification of
the cost function. This is because for an unconstrained strategy space the
effective arrival rate to the system can be larger than the service rate, and
so the fitness cannot be computed. This is a specific constraint of the
DEPU framework. Whilst it allows for emergent behaviour
to be identified in stochastic systems with complex cost or fitness
functions, it can only do so when the fitness can be readily approximated
using DES.

\subsection{Application to the general jockeying
model}\label{sec:abacws_example}

Having verified that DEPU recovers known analytical results, we now apply
it to the general jockeying model of Section~\ref{sec:general_model},
for which no closed-form fitness expressions are available. We first
extend the strategy space of the two-node example of
Section~\ref{sec:des} to demonstrate DEPU on a richer competition and
 we then move to a more complex four-floor system.

We extend the two-node
strategy space of Equation~\eqref{equ:small_example} to six strategies,
formed by combining the two visit orders \((1, 2)\) and \((2, 1)\) with
three jockeying-time levels \(j \in \{1, 5, 25\}\). The remaining model
parameters match Section~\ref{sec:des}: \(\mu = (0.5, 0.5)\),
\(c = (20, 10)\), \(\beta = (0, 0, 10)\), \(\kappa = 0.2\); DERD uses
learning rate \(\alpha = 0.01\) and step size \(\Delta t = 0.001\).
The total service capacity is \(\sum_i \mu_i c_i = 15\), so demand
values above 15 push the system into the heavy-loss regime.

\begin{figure}[!hbtp]
    \centering
    \includegraphics[width=\textwidth]{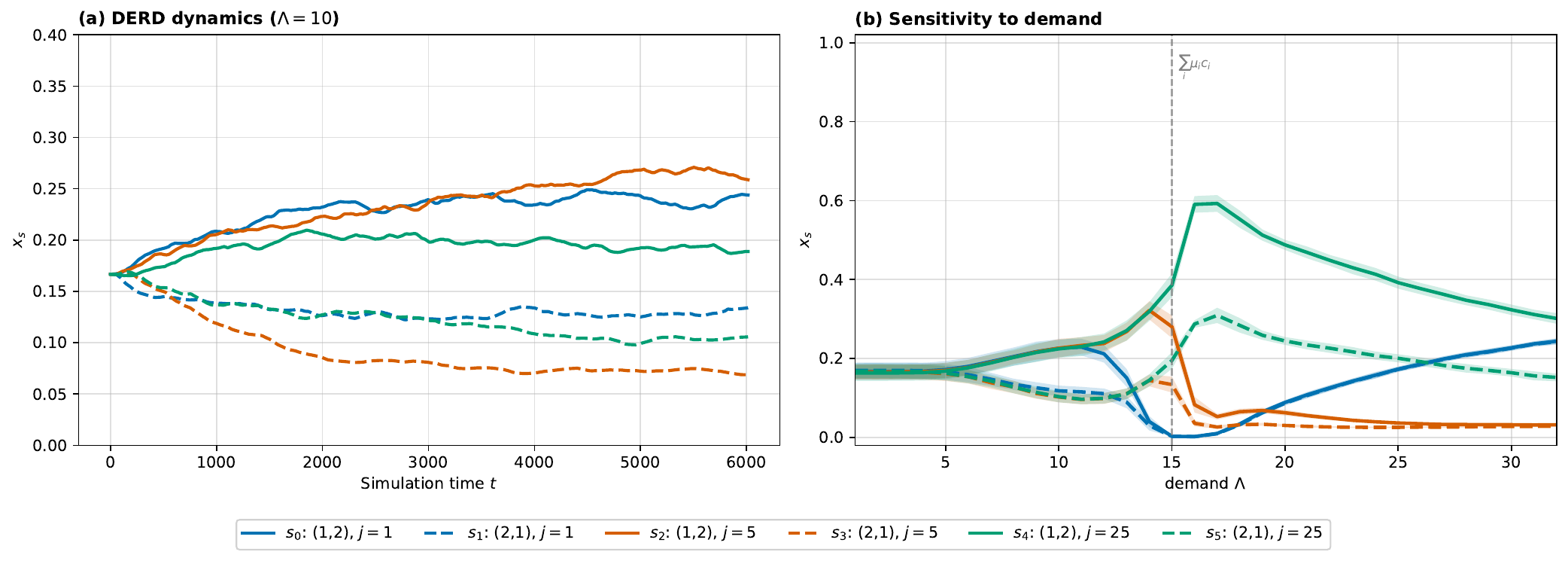}
    \caption{DERD applied to the two-node model with the extended
    six-strategy space (\(\mu = (0.5, 0.5)\), \(c = (20, 10)\),
    \(\beta = (0, 0, 10)\), \(\kappa = 0.2\), \(\alpha = 0.01\),
    \(\Delta t = 0.001\)). \textbf{(a)} Dynamics at \(\Lambda = 10\):
    population proportions \(x_s\) over simulation time, starting from
    an equal initial distribution. \textbf{(b)} Stationary
    distribution as a function of demand \(\Lambda\); each line is the
    cross-seed mean of \(x_s\) over the final 30\% of a
    40{,}000-customer DERD run computed across one hundred independent seeds,
    and the shaded band shows \(\pm\) one standard deviation. The
    equilibrium reorganises sharply as \(\Lambda\) approaches the
    service capacity \(\sum_i \mu_i c_i = 15\): patient strategies
    take over, with \(s_4\) (visit order \((1, 2)\), \(j = 25\))
    dominant from \(\Lambda \approx 17\) and decaying gradually at
    higher demand.}
    \label{fig:two_node_extended_strategy_space}
\end{figure}

Figure~\ref{fig:two_node_extended_strategy_space}~(a) shows the DERD
dynamics at the baseline demand \(\Lambda = 10\), starting from an equal
initial distribution over the six strategies. The system is below the
service capacity of 15 and the dynamics converge to a mixed equilibrium
in which all strategies coexist. The visible split is by visit order: the
three strategies that visit the higher-capacity node first, \(s_0\),
\(s_2\) and \(s_4\) (order \((1, 2)\)), settle at larger shares (roughly
\(0.19\)--\(0.26\)) than the three that visit the lower-capacity node
first, \(s_1\), \(s_3\) and \(s_5\) (order \((2, 1)\)), which drift to
roughly \(0.07\)--\(0.13\). In other words, prioritising the higher-capacity
node (here node~1, with \(c_1 = 20\) servers against \(c_2 = 10\)) is
slightly favoured, while the jockeying-time level has little effect. No
single visit order or jockeying-time level achieves dominance, a finding
that is confirmed by panel~(b).
Figure~\ref{fig:two_node_extended_strategy_space}~(b) shows the
stationary distribution as \(\Lambda\) is swept from 1 to 32 with one hundred
independent seeds each. In the under-loaded regime (\(\Lambda < 12\))
the six strategies coexist at comparable shares between roughly 0.10
and 0.25: queueing is light enough that visit order and jockeying-time
tolerance have little effect on cost. As \(\Lambda\) approaches the
service capacity the equilibrium reorganises sharply. The
medium-patience strategies \(s_2\) and \(s_3\) (\(j = 5\)) peak
briefly around \(\Lambda = 14\) before being displaced by the
high-patience strategies. From \(\Lambda \approx 17\) onwards the
patient \((1, 2)\) strategy \(s_4\) becomes dominant, peaking at
\(x_{s_4} = 0.60\) and decaying slowly as further demand growth erodes
the gap between strategies. At high demand (\(\Lambda \ge 25\)) the two
impatient strategies \(s_0\) and \(s_1\) (\(j = 1\)) re-emerge together,
at near-identical shares regardless of visit order: at \(\Lambda = 32\)
both sit at roughly \(0.24\). When service is so heavily congested that
the chance of being served within any reasonable wait is small,
abandoning the current node quickly is a viable alternative to waiting
patiently, and it makes little difference which node one waits at; visit
order ceases to matter. The shaded bands
are narrow throughout, indicating that the equilibria are robust to
seed choice.

We now consider a more complex
instance: a four-floor study space in which individuals search for a
free seat by visiting floors in a chosen order and jockeying to the next
floor after a limited wait.

The four floors have service rates \(\mu_1 = 4\), \(\mu_2 = 7\),
\(\mu_3 = 10\), \(\mu_4 = 14\) and server counts \(c_1 = 1\), \(c_2 = 2\),
\(c_3 = 2\), \(c_4 = 1\) respectively, with individuals arriving at rate
\(\Lambda = 40\). Being lost to the system incurs a cost of \(\beta = 5\);
receiving service incurs no additional exit cost. DERD uses learning
rate \(\alpha = 0.001\) and step size \(\Delta t = 0.005\); DEMR uses
population size \(M = 500\). The selection intensity is \(\kappa = 0.2\)
throughout.

Six strategies compete, combining three visit orders with two
jockeying-time tolerances:
\begin{itemize}
    \item \(s_0\): visit floors \(1 \to 2 \to 3 \to 4\), \(j = (1,1,1,1)\);
    \item \(s_1\): visit floors \(1 \to 2 \to 3 \to 4\), \(j = (3,3,3,3)\);
    \item \(s_2\): visit floors \(4 \to 3 \to 2 \to 1\), \(j = (1,1,1,1)\);
    \item \(s_3\): visit floors \(4 \to 3 \to 2 \to 1\), \(j = (3,3,3,3)\);
    \item \(s_4\): visit floors \(2 \to 4 \to 1 \to 3\), \(j = (1,1,1,1)\);
    \item \(s_5\): visit floors \(2 \to 4 \to 1 \to 3\), \(j = (3,3,3,3)\).
\end{itemize}
Strategies \(s_0\) and \(s_1\) search upward from the ground floor;
\(s_2\) and \(s_3\) search downward from the fastest top floor; \(s_4\)
and \(s_5\) use a skip-floor order that visits the second floor first,
then the top, then the ground, then the third. Within each order, the
standard variant (\(j=1\)) moves on after one time unit, while the
patient variant (\(j=3\)) waits three times as long before jockeying.

\begin{figure}[!hbtp]
    \centering
    \includegraphics[width=\textwidth]{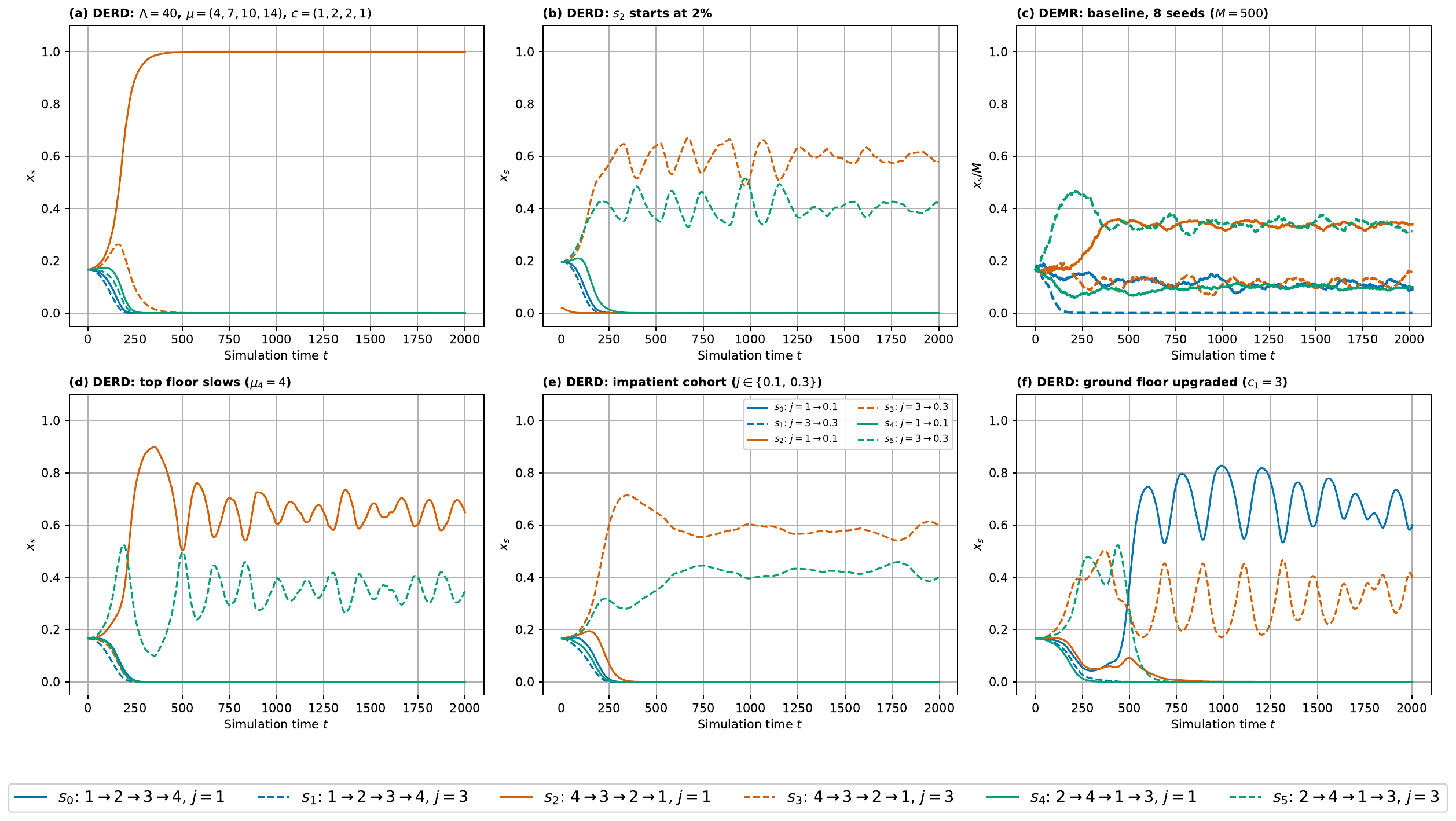}
    \caption{DEPU applied to the four-floor jockeying model
    (\(\Lambda=40\), \(\kappa=0.2\), \(\alpha=0.001\), \(c=(1,2,2,1)\)
    unless stated).
    \textbf{(a)} Baseline (\(\mu=(4,7,10,14)\)): \(s_2\) (top-down,
    \(j=1\)) is the unique ESS.
    \textbf{(b)} Initial-condition sensitivity (\(s_2\) starts at 2\%,
    others share 98\%): \(s_2\) is eliminated; \(s_3\) and \(s_5\)
    settle into an oscillating mixture, revealing multiple equilibria.
    \textbf{(c)} DEMR baseline, eight seeds (\(M=500\)): individual seeds
    fixate on different equilibria, and the cross-seed mean is dominated
    by \(s_2\) and \(s_5\), consistent with the multiple-equilibria
    finding of panel~(b).
    \textbf{(d)} Top floor slowed (\(\mu_4=4\)): oscillating mixed
    equilibrium between \(s_2\) (top-down, \(j=1\)) and \(s_5\)
    (skip-floor, \(j=3\)).
    \textbf{(e)} Impatient cohort (\(j \in \{0.1,0.3\}\)): oscillating
    mixed equilibrium between \(s_3\) (top-down, \(j=0.3\)) and \(s_5\)
    (skip-floor, \(j=0.3\)); minimal patience is decisive.
    \textbf{(f)} Ground floor refurbished (\(c_1=3\)): markedly oscillatory
    mixed equilibrium between \(s_0\) (ground-up, \(j=1\)) and \(s_3\)
    (top-down, \(j=3\)); adding servers to the ground floor displaces \(s_2\).}
    \label{fig:abacws_example}
\end{figure}

Figure~\ref{fig:abacws_example} shows the emergent behaviour under six
scenarios, organised so that the first row presents the base case, an
initial-condition sensitivity test, and the DEMR counterpart, and the
second row presents three what-if scenarios.

The six panels are as follows.

\begin{itemize}
    \item \textbf{(a) Baseline.} From an equal initial distribution DERD
        converges rapidly to \(s_2\), the top-down standard strategy, by
        around iteration 300. All other strategies are driven to zero,
        confirming \(s_2\) as an ESS for this parameter configuration.
    \item \textbf{(b) Initial-condition sensitivity.} Here \(s_2\) begins
        at just 2\% of the population while the five competitors share the
        remaining 98\% equally. Rather than climbing back to dominate,
        \(s_2\) is eliminated, and the dynamics settle into a persistent
        oscillating mixture of \(s_3\) (top-down patient, \(j=3\)) and
        \(s_5\) (skip-floor patient, \(j=3\)) at approximately 60\% and
        40\% respectively. The \(s_2\) ESS of panel~(a) is therefore not
        globally attracting: the system admits multiple equilibria, and a
        different mixed equilibrium is accessible from initial conditions
        in which the patient strategies are already well represented.
    \item \textbf{(c) DEMR baseline.} The cross-seed mean of DEMR on the
        baseline system, over eight independent seeds (\(M = 500\)) from
        approximately equal counts. The smaller population makes DEMR
        sensitive to stochastic drift, and individual seeds fixate on
        different basins: some end with \(s_2\) alone, others with \(s_5\)
        alone, and others with mixtures involving \(s_3\) or \(s_0\). The
        cross-seed mean is dominated by \(s_2\) and \(s_5\) at roughly one
        third each, consistent with panel~(b): the system supports several
        stable equilibria, and DEMR samples among them rather than picking
        one deterministically.
    \item \textbf{(d) Top floor slowed} (\(\mu_4 = 4\)). The speed
        advantage of starting at the top is removed, and an oscillating
        mixed equilibrium emerges between \(s_2\) (top-down standard,
        roughly two-thirds) and \(s_5\) (skip-floor patient, roughly
        one-third).
    \item \textbf{(e) Impatient cohort} (\(j \in \{0.1, 0.3\}\)). The same
        three visit orders compete with shorter jockeying times. The
        dynamics settle into an oscillating mixture of \(s_3\) (top-down,
        \(j=0.3\)) at roughly 60\% and \(s_5\) (skip-floor, \(j=0.3\)) at
        roughly 40\%, with the ground-up order entirely eliminated. The
        more patient of the two impatient variants wins: at such short
        jockeying times, moving on after only 0.1 time units provides
        almost no chance of service, so the slight extra patience of
        \(j=0.3\) is decisive.
    \item \textbf{(f) Ground floor refurbished} (\(c_1 = 3\)). With three
        servers the ground floor's effective throughput rises from 4 to
        12, making it a viable first stop. The dynamics settle into a
        markedly oscillatory mixed equilibrium between \(s_0\) (ground-up
        standard) and \(s_3\) (top-down patient, \(j=3\)), with \(s_0\)
        holding a majority share on average. The large oscillation
        amplitude reflects a near-neutral balance between the two
        strategies, and the baseline ESS \(s_2\) is displaced entirely.
\end{itemize}

A central advantage of DEPU over
classical RD or Moran simulation is that parameter sweeps are
tractable: each point requires only a single DES run. The dynamic
trajectories of Figure~\ref{fig:abacws_example} show how the
four-floor system responds to point changes in a single parameter; we
now sweep two control parameters to reveal how the stationary
equilibrium varies across a range of operating conditions. The swept
parameters are the demand \(\Lambda\) and a multiplicative patience
factor \(\eta\) applied uniformly to the baseline jockeying times.
Larger \(\eta\) means longer maximum waits at each floor before
jockeying. Under \(\eta\), strategies \(s_0, s_2, s_4\) have
\(j = \eta\) and strategies \(s_1, s_3, s_5\) have \(j = 3\eta\), so
that \(\eta = 1\) recovers the baseline strategy space and
\(\eta = 0.1\) recovers the impatient cohort of
Figure~\ref{fig:abacws_example}~(e). For each parameter value we run
DERD with one hundred independent seeds for 40{,}000 customers.

\begin{figure}[!hbtp]
    \centering
    \includegraphics[width=\textwidth]{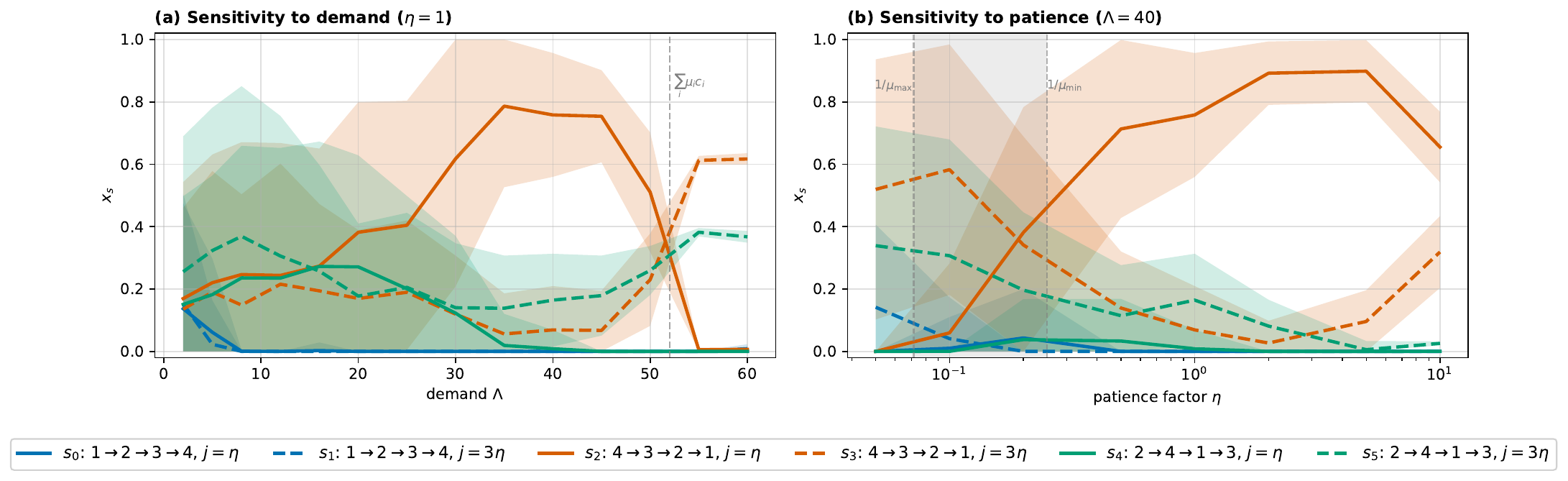}
    \caption{Stationary distribution of the four-floor jockeying model
    under two parameter sweeps. Each line is the cross-seed mean of
    \(x_s\) over the final 30\% of a 40{,}000-customer DERD run,
    averaged across one hundred independent seeds starting from an equal
    initial distribution; the shaded band shows \(\pm\) one standard
    deviation across seeds. \textbf{(a)} Sweep over demand
    \(\Lambda\) with \(\eta = 1\): the top-down standard strategy
    \(s_2\) dominates through the under-loaded and near-capacity range,
    peaking around \(\Lambda = 35\), then collapses beyond the service
    capacity, where the patient strategies \(s_3\) and \(s_5\) take
    over; the ground-up variants are eliminated except at the lightest
    demand; the dashed vertical line marks the service capacity
    \(\sum_i \mu_i c_i = 52\).
    \textbf{(b)} Sweep over the patience factor \(\eta\) with
    \(\Lambda = 40\), on a logarithmic axis: the equilibrium shifts
    from the patient variants \(s_3\) and \(s_5\) at \(\eta \le 0.1\),
    through a regime dominated by \(s_2\) for \(\eta \in [0.5, 5]\),
    and back to a mixture of \(s_2\) and \(s_3\) at \(\eta = 10\). The
    shaded vertical band spans \([1/\mu_{\max}, 1/\mu_{\min}]\), the
    range of mean service times across floors; the transition from
    patient-dominated to standard-dominated equilibrium occurs within
    this band.}
    \label{fig:abacws_stationary_sweep}
\end{figure}

Figure~\ref{fig:abacws_stationary_sweep}~(a) shows the demand sweep
spanning \(\Lambda \in [2, 60]\). The total service capacity of the
four-floor system is \(\sum_i \mu_i c_i = 52\), marked with a dashed
vertical line; the sweep therefore traverses the under-loaded,
near-capacity and over-loaded regimes. The ground-up visit orders
\(s_0\) and \(s_1\) carry appreciable mass only at the lightest demand
(\(\Lambda \le 5\)) and are eliminated thereafter, confirming that
searching from the slowest floor first is dominated by the alternative
orders. Through the under-loaded and near-capacity range the top-down
standard strategy \(s_2\) grows to dominate, peaking at roughly
four-fifths of the population around \(\Lambda = 35\). Beyond the
service capacity the equilibrium reorganises sharply: as \(\Lambda\)
passes 52, \(s_2\) collapses and is replaced by the patient top-down
strategy \(s_3\) (rising to roughly \(0.62\)) together with the patient
skip-floor strategy \(s_5\) (roughly \(0.38\)). The more patient
variants take over precisely when the system becomes overloaded,
mirroring the patience-driven transition examined next. The shaded
bands are wide through the mid-range, reflecting the multiple-equilibria
phenomenon already isolated in Figure~\ref{fig:abacws_example}~(b).
Figure~\ref{fig:abacws_stationary_sweep}~(b) shows the patience sweep
on a logarithmic axis. The shaded vertical band marks the range
\([1/\mu_{\max}, 1/\mu_{\min}] = [0.071, 0.25]\) of mean service times
across the four floors. To the left of the band, the standard
variant's maximum wait \(j = \eta\) is shorter than the mean service
time at every floor, so standard variants leave before being served and
the more patient variants \(s_3\) and \(s_5\) (with \(j = 3\eta\))
dominate the equilibrium. Inside the
band, the standard variant is served at the fastest floors but
jockeyed at the slowest, and the equilibrium transitions; the share
of the top-down standard strategy \(s_2\) rises from essentially zero
at \(\eta = 0.05\) to roughly \(0.43\) at \(\eta = 0.2\). To the right
of the band, the standard variant outlasts the typical service time
at every floor: \(s_2\) dominates and peaks near \(0.98\) at
\(\eta = 5\). At \(\eta = 10\) the patient variant \(s_3\) recovers a
non-trivial share, as further patience yields diminishing returns.
The shaded \(\pm\)\,one standard deviation bands around each strategy
are tighter than in the demand sweep and contract further at higher
\(\eta\). Taken together the two sweeps make explicit that the
equilibrium is governed jointly by visit order, jockeying-time
tolerance and demand, and that no single strategy is optimal across
the full parameter range.

\subsection{Computational efficiency}\label{sec:performance}

DEPU was motivated by the observation that the nested iterative structure of
classical RD and Moran simulation is wasteful: the inner DES is re-run from
scratch on every outer step, even though the population vector changes only
slightly between iterations. DEPU avoids this by sharing a single long DES
run between fitness estimation and population updates, and so we expect a
substantial reduction in the number of simulated customers needed to reach
a given precision. The natural question is how large the reduction is in
practice, which we now quantify.

We compare the computational cost of RD and DERD on the
jockeying example of
Section~\ref{sec:model_with_travel_times}.  We measure convergence via
the magnitude of the replicator-dynamics derivative,
\(\max_i |x_i(f_i - \bar{f})|\), where \(\bar{f} = \sum_i x_i f_i\) is
the mean fitness.  This quantity goes to zero at any evolutionary equilibrium and
does not require knowing the equilibrium analytically.  For RD, the total
customers simulated is the number of Euler iterations multiplied by the
inner DES batch size; for DERD it is simply the number of customers
processed in the single long run.

\begin{figure}[!hbtp]
\centering
\includegraphics[width=\textwidth]{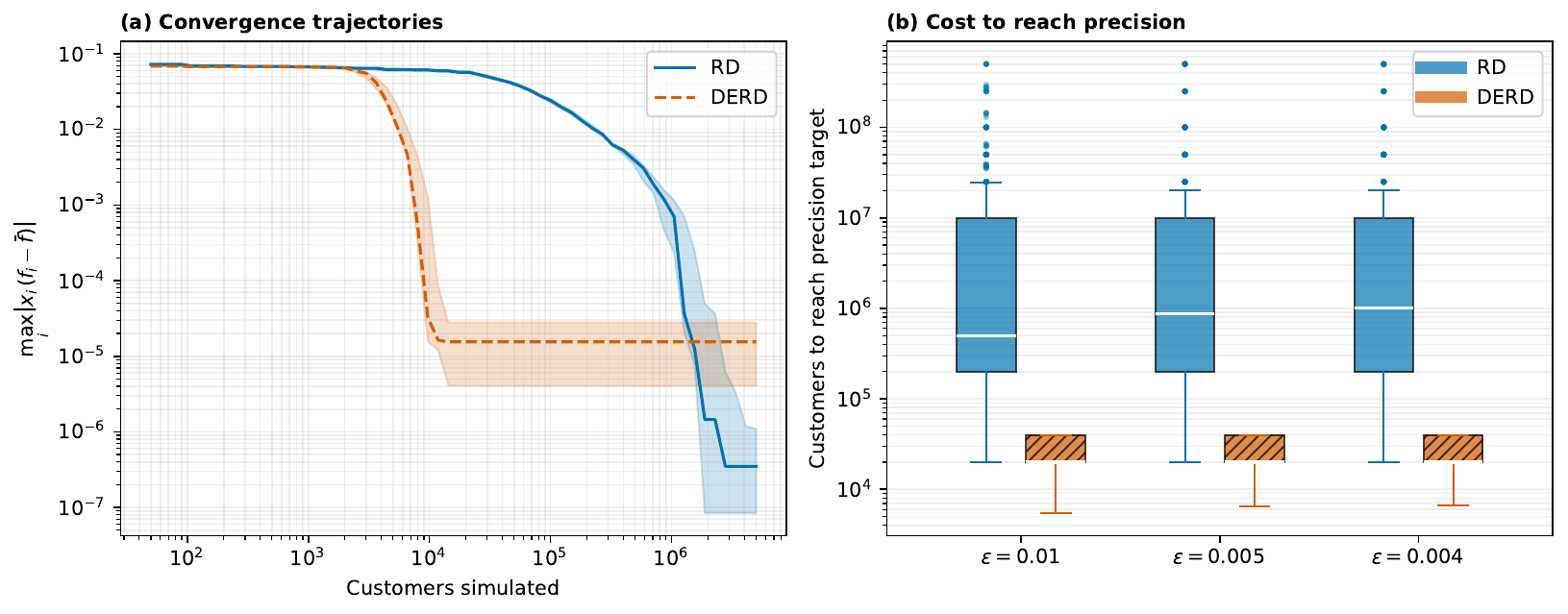}
\caption{Computational efficiency of RD (blue) and DERD (red) on the
jockeying example, measured by \(\max_i |x_i(f_i - \bar{f})|\).
(a) shows the running-minimum convergence band (median with
10th--90th percentile) across all runs and seeds; (b) shows box
plots of the customer counts needed to reach each precision target
\(\varepsilon\).}
\label{fig:performance_comparison}
\end{figure}

For each method we sweep over a grid of hyperparameters with five random
seeds per configuration. For RD we vary the inner-DES batch size
(\(\{1, 2, 5, 10, 100, 150, 250, 500\}\)), the number of Euler iterations
(\(\{20{,}000, 100{,}000\}\)), the Euler step size \(\Delta t\)
(\(\{10^{-2}, 10^{-4}, 10^{-6}\}\)) and the interval between fitness
evaluations (\(\{50, 250\}\)). For DERD we vary the total number of
customers (\(\{20{,}000, 40{,}000\}\)), the step size
\(\Delta t \in \{10^{-3}, 10^{-5}, 10^{-7}\}\), the learning rate
\(\alpha \in \{10^{-3}, 10^{-4}, 10^{-6}\}\) and the same fitness
evaluation interval as for RD.  Figure~\ref{fig:performance_comparison}a shows, for each
method, the running minimum of \(\max_i |x_i(f_i - \bar{f})|\) per seed
as a function of customers simulated, with the median and a
10th--90th percentile band across seeds.
Figure~\ref{fig:performance_comparison}b shows, for each precision target
\(\varepsilon\), box plots of the customer counts at which each individual
run first reaches \(\varepsilon\) for RD and DERD respectively.

We find that DERD consistently reaches a given precision with
substantially fewer simulated customers than RD.
Figure~\ref{fig:performance_comparison}b shows that the median customer
count required by RD at \(\varepsilon = 0.005\) is
\(875{,}000\), while the corresponding DERD median is
\(20{,}000\), a factor of roughly \(44\times\).
Figure~\ref{fig:performance_comparison}a confirms that this advantage is
consistent across seeds, with the DERD convergence band lying well below
that of RD throughout.

A caveat is that the converged DERD population fluctuates stochastically
around the equilibrium rather than settling exactly on it. This is
visible in Figure~\ref{fig:performance_comparison}a, where the DERD
convergence band flattens at a noise floor of roughly \(1.5 \times
10^{-5}\) and descends no further, whereas RD continues towards zero.
Precision targets below this floor are therefore unattainable by DERD; the
targets reported here (\(\varepsilon \ge 0.004\)) all lie comfortably above
it, so the speed-up is measured in a regime where both methods converge.

\begin{table}[!hbtp]
\centering
\caption{Median customers required by RD and DERD to reach each
precision target across all runs, with the implied speedup factor.}
\input{performance_comparison_speedup.tex}
\label{tab:performance_comparison_speedup}
\end{table}

\section{Conclusion}
\label{sec:conclusion}

This paper introduced DEPU, a general
algorithmic framework for identifying emergent equilibrium behaviour in
stochastic systems whose fitness functions do not admit closed-form
expressions. DEPU couples a single long run of a DES
directly with an evolutionary population update rule, rather than nesting
repeated short simulations inside an outer optimisation loop, and so
achieves accurate convergence to equilibrium at substantially lower
computational cost. Two implementations were presented: Discrete Event
Replicator Dynamics (DERD), which tracks a continuous population density
vector according to an Euler discretisation of the RD
equation; and Discrete Event Moran Replacement (DEMR), which maintains a
finite discrete population updated via Moran-style copying and removal
events. Both implementations were validated against the known analytical
equilibrium of~\cite{naor1969regulation}, and then applied to a jockeying
queueing model for which no closed-form fitness expressions exist.

The behavioural queueing literature has, with few exceptions, been
confined to the small class of systems admitting analytical closed forms,
and the toolkit of evolutionary dynamics has been similarly confined.
Coupling DES to evolutionary updates removes both
restrictions at once, and brings under behavioural analysis the much
larger class of queueing systems that practitioners actually build and
operate. The framework is agnostic to the underlying simulator, the
specific revision protocol, and the particular fitness functional, so the
same architecture applies equally well outside queueing, to any
population game whose payoffs can be sampled from a DES.

Building
evacuation is one such setting: occupants choose which stairwells or
exits to attempt in sequence, and the effectiveness of any route depends
on how many others choose the same path~\cite{helbing2000simulating,
lovreglio2016review}. DES is the dominant
methodology for evacuation planning~\cite{robinson14}, so DEPU could be
applied directly to existing simulation models to identify which
evacuation strategies emerge under self-interested behaviour, rather
than assuming all occupants follow a prescribed plan. The same logic
applies to patients choosing between emergency departments with
different waiting times, or travellers choosing between routes at a
multi-lane checkpoint. In each case a DES model already exists in
practice; DEPU provides a route to the emergent strategic equilibrium
without requiring any closed-form fitness expression.

A natural direction for future work is to embed
alternative revision protocols within the DEPU framework, replacing the
population update step in Figure~\ref{fig:depu} while leaving the DES-based
fitness estimation unchanged.

\emph{Imitation dynamics} provide one such alternative. Under imitation
rules, agents periodically compare their current strategy against that of a
randomly sampled peer and switch with a probability that increases in the
fitness advantage of the peer's strategy~\cite{schlag1998}. Unlike the
RD, which posits an implicitly global averaging of fitness
across the population, imitation dynamics are grounded in local, pairwise
comparisons, and may therefore be more behaviourally realistic in settings
where individuals cannot observe system-level averages. A DEPU implementation
based on imitation would trigger a pairwise comparison event each time a
customer exits the system and receives a cost, updating the population
vector by probabilistically transferring weight from lower-fitness to
higher-fitness strategies.

\emph{Best response dynamics}~\cite{gilboamatsui1991} represent a second
alternative, in which agents occasionally switch to whichever strategy
maximises their expected fitness given the current population. In the DEPU
context, this would require estimating the best response at each update step
from the running DES fitness estimates, and then shifting population weight
deterministically towards that strategy. Best response dynamics can exhibit
more abrupt transitions than RD and may converge faster in
strongly asymmetric settings.

\emph{Introspection dynamics}~\cite{sandholm2010} offer a third direction.
Under introspection, agents revise their strategy by comparing their current
payoff against the payoff they would \emph{hypothetically} receive under a
randomly drawn alternative strategy, rather than against the realised payoff
of an observed peer. This revision protocol sits between the purely social
learning of imitation dynamics and the fully forward-looking calculation of
best response dynamics, and has been shown to select among Nash equilibria
in ways that differ from both replicator and best response dynamics.
Embedding introspection within DEPU would require maintaining hypothetical
fitness estimates for each strategy type, which could be achieved by
applying the same exponential-smoothing update rule introduced in
Section~\ref{sec:depu} to a separate set of fitness trackers.

More broadly, the modular structure of DEPU, in which fitness estimation
and population updating are decoupled, means that any revision protocol
that can be expressed as a map from a fitness vector to a population
update can in principle be substituted in. The taxonomy of such protocols
catalogued in~\cite{sandholm2010} therefore represents a collection of
potential DEPU variants, each with its own behavioural interpretation and
convergence properties, all accessible without any change to the
underlying DES. Extending DEPU in this direction,
and characterising the resulting variants, is a natural open problem.

The entire pipeline, from raw data generation
through every figure and table in this paper, is implemented to current
standards for reproducible computational research~\cite{sandve2013ten}
and follows the FAIR principles for scientific data
management~\cite{wilkinson2016fair}. Source code, parameter
configurations, and analysis scripts are version-controlled together,
dependencies are pinned via \texttt{uv}, and stage dependencies between
data and analysis are tracked with \texttt{DVC} so that every artefact
can be regenerated from a single command. The data and code are archived
at Zenodo under DOI \texttt{10.5281/zenodo.20931567}.

We hope that DEPU enables behavioural analysis of queueing systems that
have, until now, been studied only under the assumption that all
individuals follow a single strategy, and that the modular structure of
the framework will encourage adoption and adaptation across the
population-game literature more broadly.

\bibliographystyle{plain}
\bibliography{bibliography.bib}
\end{document}

%% file: jockeying_diagram.tex
\begin{tikzpicture}
  % Queue N
  \draw (0, 0) -- (8, 0);
  \draw (8, 0) -- (8, 3);
  \draw (0, 3) -- (8, 3);
  \draw (4, 0) -- (4, 3);
  \draw (5, 0) -- (5, 3);
  \draw (6, 0) -- (6, 3);
  \draw (7, 0) -- (7, 3);
  % Service centers
  \draw (10.5, 1.5) circle (1.5);
  \node at (10.5, 2) {\huge{$c_{\pi_n}$}};
  \node at (10.5, 1) {\huge{$G_{\pi_n}$}};

  % Dots
  \node at (10.5, 5.25) {\Huge{$\vdots$}};

  % Queue 1
  \draw (0, 7.5) -- (8, 7.5);
  \draw (8, 7.5) -- (8, 10.5);
  \draw (0, 10.5) -- (8, 10.5);
  \draw (4, 7.5) -- (4, 10.5);
  \draw (5, 7.5) -- (5, 10.5);
  \draw (6, 7.5) -- (6, 10.5);
  \draw (7, 7.5) -- (7, 10.5);
  % Service centers
  \draw (10.5, 9) circle (1.5);
  \node at (10.5, 9.5) {\huge{$c_{\pi_2}$}};
  \node at (10.5, 8.5) {\huge{$G_{\pi_2}$}};

  % Queue 0
  \draw (0, 12) -- (8, 12);
  \draw (8, 12) -- (8, 15);
  \draw (0, 15) -- (8, 15);
  \draw (4, 12) -- (4, 15);
  \draw (5, 12) -- (5, 15);
  \draw (6, 12) -- (6, 15);
  \draw (7, 12) -- (7, 15);
  % Service centers
  \draw (10.5, 13.5) circle (1.5);
  \node at (10.5, 14) {\huge{$c_{\pi_1}$}};
  \node at (10.5, 13) {\huge{$G_{\pi_1}$}};

  % Arrivals
  \draw[thick, -triangle 45] (-5.5, 13.5) -- (-1, 13.5);
  \node at (-5, 14) {\huge{$\Lambda$}};
  % Departures
  \draw[thick, -triangle 45] (12.5, 1.5) -- (16.5, 1.5);
  \node at (16, 2) {\huge{$\beta_{\pi_n}$}};
  \draw[thick, -triangle 45] (12.5, 9) -- (16.5, 9);
  \node at (16, 9.5) {\huge{$\beta_{\pi_2}$}};
  \draw[thick, -triangle 45] (12.5, 13.5) -- (16.5, 13.5);
  \node at (16, 14) {\huge{$\beta_{\pi_1}$}};

  % Jockeying
  \node (out_0) at (1.5, 13) {};
  \node (out_1) at (1.5, 8.5) {};
  \node (out_dots) at (2.25, 4.25) {};
  \node (out_N) at (1.5, 1) {};
  \draw[thick, -triangle 45] (out_0) edge[out=-45, in=45] (1.5, 9.5);
  \node at (2.75, 12.5) {\huge{$j_{\pi_1}$}};
  \draw[thick, -triangle 45] (out_1) edge[out=-45, in=90] (2.25, 6.25);
  \node at (2.75, 8) {\huge{$j_{\pi_2}$}};
  \draw[thick, -triangle 45] (out_dots) edge[out=-90, in=45] (1.5, 2);
  \node at (3.25, 3.75) {\huge{$j_{\pi_{n-1}}$}};
  \draw[thick, -triangle 45] (out_N) edge[out=-45, in=-180] (16.5, -2.5);
  \node at (3.25, -1.5) {\huge{$j_{\pi_{n}}$}};
  \node at (16, -2) {\huge{$\beta_{N + 1}$}};

\end{tikzpicture}

%% file: event_scheduling_diagram.tex
\begin{tikzpicture}

\node[align=center] (initial) [style={minimum size=4cm, minimum height=1.5cm, text width=2cm, draw=black,fill=black!5,text=black,shape=ellipse}] at (0, 7) {\large{Initialise Simulation}};

\node[align=center] (Aphase) [style={minimum width=3.5cm, minimum height=1cm, text width=4cm, draw=black,fill=black!5,text=black,shape=rectangle}] at (0, 4) {\large{\textbf{A}-Phase}};
\node[align=center] (Bphase) [style={minimum width=3.5cm, minimum height=1cm, text width=4cm, draw=black,fill=black!5,text=black,shape=rectangle}] at (-7, 4) {\large{\textbf{B}-Phase}};
\node[align=center] (Cphase) [style={minimum width=3.5cm, minimum height=1cm, text width=4cm, draw=black,fill=black!5,text=black,shape=rectangle}] at (-14, 4) {\large{\textbf{C}-Phase}};

\node[align=center] (anymoreC) [style={minimum size=2cm, text width=1.8cm, draw=black,fill=black!5,text=black,shape=diamond, aspect=2}] at (-14, 0) {\small{Any more \textbf{C}-Events?}};
\node[align=center] (anymoreB) [style={minimum size=2cm, text width=1.8cm, draw=black,fill=black!5,text=black,shape=diamond, aspect=2}] at (-7, 0) {\small{Any more \textbf{B}-Events?}};
\node[align=center] (isend) [style={minimum size=2cm, text width=1.8cm, draw=black,fill=black!5,text=black,shape=diamond, aspect=2}] at (0, 0) {\small{Simulation Complete?}};

\node[align=center] (stop) [style={minimum size=4cm, minimum height=1.5cm, text width=1.8cm, draw=black,fill=black!5,text=black,shape=ellipse}] at (0, -3.5) {\large{End Simulation}};

\draw[draw=none] (0, 0) -- (0, -2);

\draw[thick, -triangle 45] (initial) -- (Aphase);
\draw[thick, -triangle 45] (Aphase) -- (Bphase);
\draw[thick, -triangle 45] (Bphase) -- (Cphase);
\draw[thick, -triangle 45] (anymoreC) -- (Cphase);
\draw[thick, -triangle 45] (anymoreC) -- (anymoreB);
\draw[thick, -triangle 45] (anymoreB) -- (isend);
\draw[thick, -triangle 45] (isend) -- (stop);
\draw[thick, -triangle 45] (Cphase) -- (-18, 4) -- (-18, 0) -- (anymoreC);
\draw[thick, -triangle 45] (anymoreB) -- (Bphase);
\draw[thick, -triangle 45] (isend) -- (Aphase);

\node at (-0.4, -2) {Yes};
\node at (-3.5, 0.25) {No};
\node at (-10.5, 0.25) {No};
\node at (-0.4, 2) {No};
\node at (-7.4, 2) {Yes};
\node at (-14.4, 2) {Yes};

\end{tikzpicture}

%% file: get_fitness_by_simulation.tex
\begin{tikzpicture}

\draw[thick, dashed, draw=black, fill=black!5] (-19, -5) rectangle (3, 9);
\node at (-17, 8.25) {\LARGE{\textit{Find fitness}}};

\node[align=center] (start) [style={minimum size=4cm, minimum height=1.5cm, text width=2cm, draw=black,fill=black!5,text=black,shape=ellipse}] at (7, 10) {\large{Start}};

\node[align=center] (initial) [style={minimum width=3.5cm, minimum height=1cm, text width=4cm, draw=black,fill=black!5,text=black,shape=rectangle}] at (0, 7) {\large{Initialise Simulation}};

\node[align=center] (Aphase) [style={minimum width=3.5cm, minimum height=1cm, text width=4cm, draw=black,fill=black!5,text=black,shape=rectangle}] at (0, 4) {\large{\textbf{A}-Phase}};
\node[align=center] (Bphase) [style={minimum width=3.5cm, minimum height=1cm, text width=4cm, draw=black,fill=black!5,text=black,shape=rectangle}] at (-7, 4) {\large{\textbf{B}-Phase}};
\node[align=center] (Cphase) [style={minimum width=3.5cm, minimum height=1cm, text width=4cm, draw=black,fill=black!5,text=black,shape=rectangle}] at (-14, 4) {\large{\textbf{C}-Phase}};

\node[align=center] (anymoreC) [style={minimum size=2cm, text width=1.8cm, draw=black,fill=black!5,text=black,shape=diamond, aspect=2}] at (-14, 0) {\small{Any more \textbf{C}-Events?}};
\node[align=center] (anymoreB) [style={minimum size=2cm, text width=1.8cm, draw=black,fill=black!5,text=black,shape=diamond, aspect=2}] at (-7, 0) {\small{Any more \textbf{B}-Events?}};
\node[align=center] (isend) [style={minimum size=2cm, text width=1.8cm, draw=black,fill=black!5,text=black,shape=diamond, aspect=2}] at (0, 0) {\small{Simulation Complete?}};

\node[align=center] (endsim) [style={minimum width=3.5cm, minimum height=1cm, text width=4cm, draw=black,fill=black!5,text=black,shape=rectangle}] at (0, -3.5) {\large{End Simulation}};

\node[align=center] (stopcriteria) [style={minimum size=2cm, text width=1.8cm, draw=black,fill=black!5,text=black,shape=diamond, aspect=2}] at (7, -3.5) {\small{Stopping criteria?}};

\node[align=center] (updatepop) [style={minimum width=3.5cm, minimum height=1cm, text width=4cm, draw=black,fill=black!5,text=black,shape=rectangle}] at (7, 2) {\large{Update Population}};

\node[align=center] (stop) [style={minimum size=4cm, minimum height=1.5cm, text width=1.8cm, draw=black,fill=black!5,text=black,shape=ellipse}] at (7, -7) {\large{Stop}};

\draw[draw=none] (0, 0) -- (0, -2);

\draw[thick, -triangle 45] (start) -- (0, 10) -- (initial);
\draw[thick, -triangle 45] (initial) -- (Aphase);
\draw[thick, -triangle 45] (Aphase) -- (Bphase);
\draw[thick, -triangle 45] (Bphase) -- (Cphase);
\draw[thick, -triangle 45] (anymoreC) -- (Cphase);
\draw[thick, -triangle 45] (anymoreC) -- (anymoreB);
\draw[thick, -triangle 45] (anymoreB) -- (isend);
\draw[thick, -triangle 45] (isend) -- (endsim);
\draw[thick, -triangle 45] (Cphase) -- (-18, 4) -- (-18, 0) -- (anymoreC);
\draw[thick, -triangle 45] (anymoreB) -- (Bphase);
\draw[thick, -triangle 45] (isend) -- (Aphase);
\draw[thick, -triangle 45] (endsim) -- (stopcriteria);
\draw[thick, -triangle 45] (stopcriteria) -- (stop);
\draw[thick, -triangle 45] (stopcriteria) -- (updatepop);
\draw[thick, -triangle 45] (updatepop) -- (7, 7) -- (initial);

\node at (-0.4, -2) {Yes};
\node at (-3.5, 0.25) {No};
\node at (-10.5, 0.25) {No};
\node at (-0.4, 2) {No};
\node at (-7.4, 2) {Yes};
\node at (-14.4, 2) {Yes};
\node at (6.6, -5) {Yes};
\node at (6.6, -1.5) {No};

\end{tikzpicture}

%% file: depu_diagram.tex
\begin{tikzpicture}

\node[align=center] (start) [style={minimum size=4cm, minimum height=1.5cm, text width=2cm, draw=black,fill=black!5,text=black,shape=ellipse}] at (7, 7) {\large{Start}};

\node[align=center] (Aphase) [style={minimum width=3.5cm, minimum height=1cm, text width=4cm, draw=black,fill=black!5,text=black,shape=rectangle}] at (0, 4) {\large{\textbf{A}-Phase}};
\node[align=center] (Bphase) [style={minimum width=3.5cm, minimum height=1cm, text width=4cm, draw=black,fill=black!5,text=black,shape=rectangle}] at (-7, 4) {\large{\textbf{B}-Phase}};
\node[align=center] (Cphase) [style={minimum width=3.5cm, minimum height=1cm, text width=4cm, draw=black,fill=black!5,text=black,shape=rectangle}] at (-14, 4) {\large{\textbf{C}-Phase}};

\node[align=center] (anymoreC) [style={minimum size=2cm, text width=1.8cm, draw=black,fill=black!5,text=black,shape=diamond, aspect=2}] at (-14, 0) {\small{Any more \textbf{C}-Events?}};
\node[align=center] (anymoreB) [style={minimum size=2cm, text width=1.8cm, draw=black,fill=black!5,text=black,shape=diamond, aspect=2}] at (-7, 0) {\small{Any more \textbf{B}-Events?}};
\node[align=center] (stopcriteria) [style={minimum size=2cm, text width=1.8cm, draw=black,fill=black!5,text=black,shape=diamond, aspect=2}] at (0, 0) {\small{Stoppping criteria?}};

\node[align=center] (updatepop) [style={minimum width=3.5cm, minimum height=1cm, text width=4cm, draw=black,fill=black!5,text=black,shape=rectangle}] at (7, 2) {\large{Update Population}};

\node[align=center] (stop) [style={minimum size=4cm, minimum height=1.5cm, text width=1.8cm, draw=black,fill=black!5,text=black,shape=ellipse}] at (7, -3) {\large{Stop}};

\draw[draw=none] (0, 0) -- (0, -2);

\draw[thick, -triangle 45] (start) -- (0, 7) -- (Aphase);
\draw[thick, -triangle 45] (Aphase) -- (Bphase);
\draw[thick, -triangle 45] (Bphase) -- (Cphase);
\draw[thick, -triangle 45] (anymoreC) -- (Cphase);
\draw[thick, -triangle 45] (anymoreC) -- (anymoreB);
\draw[thick, -triangle 45] (anymoreB) -- (stopcriteria);
\draw[thick, -triangle 45] (Cphase) -- (-18, 4) -- (-18, 0) -- (anymoreC);
\draw[thick, -triangle 45] (anymoreB) -- (Bphase);
\draw[thick, -triangle 45] (stopcriteria) -- (0, -3) -- (stop);
\draw[thick, -triangle 45] (stopcriteria) -- (7, 0) -- (updatepop);
\draw[thick, -triangle 45] (updatepop) -- (7, 4) -- (Aphase);

\node at (-0.4, -2) {Yes};
\node at (-3.5, 0.25) {No};
\node at (-10.5, 0.25) {No};
\node at (-7.4, 2) {Yes};
\node at (-14.4, 2) {Yes};
\node at (3, 0.25) {No};

\end{tikzpicture}

%% file: compare_depu_rd_mp.tex
\begin{tikzpicture}

\node[align=center] (DEPU) at (0, 6.15) {\it DEPU};
\draw[draw=black, rounded corners, fill=black!5!white] (-1.25, -1) rectangle (1.25, 5.85);
\node[align=center, minimum width=2cm, rounded corners, draw=black, fill=black!10!white, outer sep=0.4cm] (DEMR) at (0, 0) {\large{DEMR}};
\node[align=center, minimum width=2cm, rounded corners, draw=black, fill=black!10!white, outer sep=0.4cm] (DERP) at (0, 5) {\large{DEMR}};
\node[align=center, minimum width=2cm, rounded corners, draw=black, fill=black!10!white, outer sep=0.4cm] (RD) at (7, 5) {\large{RD}};
\node[align=center, minimum width=2cm, rounded corners, draw=black, fill=black!10!white, outer sep=0.4cm] (MP) at (7, 0) {\large{MP}};

\draw [triangle 90 - triangle 90] (RD) -- (DERP);
\draw [triangle 90 - triangle 90] (RD) -- (MP);
\draw [triangle 90 - triangle 90] (DERP) -- (DEMR);

\node at (3.5, 5.15) {\footnotesize Equivalent};
\node at (3.5, 4.85) {\footnotesize in expectation};
\node[rotate=90] at (6.85, 2.5) {\footnotesize Equivalent as};
\node[rotate=90] at (7.18, 2.5) {\footnotesize $M \rightarrow \infty$ (\cite{traulsen2005coevolutionary})};
\node[rotate=90] at (-0.15, 2.5) {\footnotesize Equivalent as};
\node[rotate=90] at (0.15, 2.5) {\footnotesize $M \rightarrow \infty$ and $t \rightarrow \infty$};

\end{tikzpicture}

%% file: naor_diagram.tex
\begin{tikzpicture}
  % Queue 0
  \draw (0, 12) -- (8, 12);
  \draw (8, 12) -- (8, 15);
  \draw (0, 15) -- (8, 15);
  \draw (4, 12) -- (4, 15);
  \draw (5, 12) -- (5, 15);
  \draw (6, 12) -- (6, 15);
  \draw (7, 12) -- (7, 15);
  % Service centers
  \draw (10.5, 13.5) circle (1.5);
  \node at (10.5, 13.5) {\huge{$\mu$}};

  % Arrivals
  \draw[thick, -triangle 45] (-5.5, 13.5) -- (-1, 13.5);
  \node at (-5, 14) {\huge{$\Lambda$}};
  % Departures
  \draw[thick, -triangle 45] (12.5, 13.5) -- (15, 13.5);
  \node at (16, 14) {\huge{$T$}};

  % Jockeying
  \node (out_0) at (1.5, 13) {};
  \draw[thick, -triangle 45] (out_0) edge[out=-45, in=45] (1.5, 9.5);
  \node at (2.75, 12.5) {\huge{$\beta$}};

\end{tikzpicture}

%% file: performance_comparison_speedup.tex
\begin{tabular}{lrrrr}
\toprule
Precision $\varepsilon$ & Median RD customers & Median DERD customers & Median speedup \\
\midrule
$0.01$ & $500,000$ & $20,000$ & $25\times$ \\
$0.005$ & $875,000$ & $20,000$ & $44\times$ \\
$0.004$ & $1,000,000$ & $20,000$ & $50\times$ \\
\bottomrule
\end{tabular}